\documentclass[reprint,preprintnumbers,
amsmath,amssymb,aps,nofootinbib,superscriptaddress]{revtex4-1}
\usepackage[pdftex]{graphicx}
\usepackage{dcolumn}
\usepackage{bm}
\usepackage{natbib}
\usepackage{cases}
\usepackage[pdftex]{hyperref}
\usepackage{color}
\usepackage{subfigure}
\makeatletter \def\@eqnnum{{\normalsize \normalcolor (\theequation)}}  
\makeatother %
\newcommand{\para}[1]{\par\vspace{2mm}\noindent\textbf{\emph{{#1}}.---}}
\definecolor{darkblue}{rgb}{0.15,0.35,0.55}
\definecolor{reddish}{rgb}{0.65, 0.2, 0.2}
\definecolor{indigo(dye)}{rgb}{0.0, 0.25, 0.42}
\hypersetup{colorlinks=true,linkcolor=reddish,
citecolor=indigo(dye),filecolor=indigo(dye),urlcolor=indigo(dye)}
\usepackage{comment}
\usepackage[utf8]{inputenc}

\begin{document}
\preprint{\begin{minipage}[b]{1\linewidth} 
\begin{flushright}Preprint numbers: TU-1092\end{flushright}
\end{minipage}}
\title{Quantum Spacetime Instability and Breakdown of Semiclassical Gravity  }
\author{Hiroki Matsui}
\email{hiroki.matsui.c6@tohoku.ac.jp}
\affiliation{Department of Physics, Tohoku University, 
Sendai, Miyagi 980-8578, Japan}
\author{Naoki Watamura}
\email{watamura@shu.edu.cn}
\affiliation{Department of Mathematics, Shanghai University, Shanghai, China}

\begin{abstract}
The semiclassical gravity describes gravitational
back-reactions of the classical spacetime interacting with quantum matter fields 
but the quantum effects on the background is formally defined as
higher derivative curvatures. These induce catastrophic instabilities 
and classic solutions become unstable under small perturbations or their evolutions. 
In this paper we discuss validity of the semiclassical gravity
from the perspective of the spacetime instabilities and 
consider cosmological dynamics of the Universe in this theory.
We clearly show that the homogenous and isotropic flat Universe is unstable 
and the solutions either grow exponentially 
or oscillate even in Planckian time $t_{\rm I}=(\alpha_{ 1 }G_N)^{1/2}\approx 
\alpha_{ 1 }10^{-43}\ {\rm sec}$. 
The subsequent curvature evolution leads to Planck-scale spacetime curvature 
in a short time and causes a catastrophe of the Universe 
unless one takes extremely large values of the gravitational couplings. 
Furthermore, we confirm the above suggestion by comparing 
the semiclassical solutions and $\Lambda$CDM with the Planck data 
and it is found that
the semiclassical solutions are not consistent with the cosmological observations.
Thus, the standard semiclassical gravity using 
quantum energy momentum tensor $\left< { T }_{ \mu \nu  } \right>$
is not appropriate to describe our Universe. 
\end{abstract}
\date{\today}
\maketitle
\flushbottom
\allowdisplaybreaks[0]

\section{Introduction}
\label{sec:intro}
There are many essential difficulties to construct 
a consistent theory of quantum gravity.
Almost certainly, quantum gravity where metric is also quantized together with matter fields
would change even fundamental concept 
of the spacetime and requires a completely different theory from classical general relativity.
However, in the regime where the curvature is small,
one usually regard the gravity as a classic field and matters 
move on the background. Therefore, 
the semiclassical approximation is usually expected to be sufficient. 
Based on this assumption,
thermal Hawking radiation around black holes~\cite{Hawking:1974sw} or
amplification of primordial quantum fluctuations during inflation 
are correctly performed by quantum field theory in curved spacetime~\cite{Birrell:1982ix}.

However, not only quantum fluctuations of the matter fields, but also 
quantum back-reaction on the spacetime must be considered 
in full semiclassical approximation.
The standard semiclassical gravity
replaces energy momentum tensor 
in Einstein equations by the expectation values of some quantum state,
\begin{align}\label{eq:semiclassical}
{ G }_{ \mu \nu  } +{ \Lambda  } { g }_{ \mu\nu }
\equiv8\pi { G }_{ N }\left< {\Psi  }|{ T }_{ \mu\nu }|
{ \Psi } \right>.
\end{align}
The semiclassical gravity naturally includes
quantum effects of the matters on spacetime 
such as vacuum polarization or quantum particle creation.
This theory is regarded as a first-approximation to quantum gravity~\cite{Buchbinder:1992rb} 
and yields some insight into quantum nature of gravity.
For instance, it describes evaporation 
of the black holes~\cite{Hawking:1974sw,Unruh:1976db,
Christensen:1977jc,Candelas:1980zt,Ho:2018fwq} and
provides new classes of the cosmological solutions~\cite{Davies:1977ti,Fischetti:1979ue,
Starobinsky:1980te,Anderson:1983nq,Anderson:1984jf,Azuma:1986st,Nojiri:2003vn,Nojiri:2004ip}.
However, the quantum energy momentum tensor 
has non-trivial structures which depend on the curvature tensor or its derivatives
and introduce higher derivative corrections.
In the semiclassical gravity these higher derivative curvatures always appear 
and are necessary to take account of the interaction of the classical gravitational field 
with quantum matter fields.
Even in quantum gravity the higher derivative curvatures 
are necessary for the renormalizability of the theory~\cite{Utiyama:1962sn}. 
However, these induce instabilities of classical spacetime 
drawn from general relativity and produce unphysical massive ghosts
which leads to the non-unitary graviton 
$S$-matrix~\cite{Stelle:1976gc,Salles:2014rua}.
Although we usually assume that the semiclassical gravity would be 
applicable below the Planck regime, it has several undesired properties 
and the validity is still unknown.

In fact, Refs~\cite{Horowitz:1978fq,Horowitz:1980fj,Hartle:1981zt,
RandjbarDaemi:1981wd,Jordan:1987wd,Suen:1988uf,Suen:1989bg,Anderson:2002fk}
have shown that the Minkowski spacetime in the semiclassical gravity
is unstable under small perturbations and it is 
not the ground state~\cite{Hartle:1981zt}.
The perturbations either grow exponentially 
or oscillate even in the Planck time
and the subsequent curvature evolution 
leads to the Planck-scale spacetime curvature~\cite{Horowitz:1978fq,Horowitz:1980fj}.
By using large $N$ expansion of quantum gravity~\cite{Hartle:1981zt} or 
effective action approach~\cite{Jordan:1987wd} 
it was shown that the curvature instabilities occur at the frequencies far below the Planck regime. 
The quantum de Sitter instability for scalar fields or graviton
has been also discussed by Refs~\cite{Ford:1984hs,Mottola:1984ar,Mottola:1985qt,Antoniadis:1985pj,Antoniadis:1986sb,Higuchi:1986ww,Polarski:1990tr,Tsamis:1992sx,Tsamis:1994ca,
Tsamis:1992xa,Tsamis:1996qq,Tsamis:1996qk,Mukhanov:1996ak,Abramo:1997hu,Goheer:2002vf,Brandenberger:2002sk,Kachru:2003aw,Finelli:2004bm,Janssen:2007ht,Janssen:2008dw,Janssen:2008dp,Polyakov:2009nq,Shukla:2016bnu,Anderson:2013ila,Anderson:2013zia,
Myrzakulov:2014hca,Cusin:2015oza,Salles:2017xsr,Kuntz:2019gup}.
These facts strongly indicate that the semiclassical gravity 
is not a good theory of the gravity and the instabilities
cannot be easily rescued by full 
quantum theory of gravity~\cite{Hartle:1981zt,Jordan:1987wd}.
However, it has not been clearly shown whether the quantum instabilities are incompatible with
the cosmological observations.

In this paper, we will thoroughly investigate the spacetime instabilities
induced by quantum back-reaction and 
reconsider cosmological dynamics of the Universe in the
semiclassical equations.
Due to the spacetime instabilities,
we obtain exactly nontrivial
cosmological constraints on the semiclassical gravity.
We clearly show that Minkowski spacetime or 
homogenous and isotropic flat spacetime are unstable 
and the corresponding solutions either grow exponentially 
or oscillate even in the Planckian time $t_{\rm I}=(\alpha_{ 1 }G_N)^{1/2}\approx 
\alpha_{ 1 }10^{-43}\ {\rm sec}$ where $\alpha_{ 1 }$ is defined by the 
quantum energy momentum tensor or higher derivative gravitational action.
Furthermore, we confirm the above proposition by comparing 
the cosmological solutions of the $\Lambda$CDM and 
the semiclassical Einstein equations with the recent Planck data~\cite{Aghanim:2018eyx}
and then, we show that the semiclassical gravity is inconsistent with  
the cosmological observations unless one takes extremal values 
of the gravitational couplings. Thus, the standard semiclassical gravity using 
quantum energy momentum tensor is not appropriate to describe our Universe.

The present paper is organized as follows. 
In Section~\ref{sec:semiclassical}, we review the semiclassical gravity
and introduce our formulation for this theory.
In particular, we explain how the higher-derivative corrections  
appear in the semiclassical gravity. 
Furthermore, we discuss several problems of the semiclassical gravity
such as violations of gravitational thermodynamical laws and  
(averaged) null energy condition. 
In Section~\ref{sec:instability}, 
we investigate the spacetime instabilities 
induced by quantum back-reaction.
First, we consider some results of Refs~\cite{Horowitz:1978fq,
Suen:1988uf,Suen:1989bg} and 
investigate the instability of Minkowski spacetime 
under small perturbations.
Next, we investigate quantum instabilities of the homogenous and 
isotropic FLRW Universe and obtain cosmological constraints on the 
semiclassical gravity.
Finally, in Section~\ref{sec:discussion} we discuss the validity of 
this theory and draw the conclusion of our work.

\section{Semiclassical Gravity}
\label{sec:semiclassical}
In this section we review how quantum matter fields interacts with 
the spacetime and introduce our formulation for the semiclassical gravity.
At the quantum level the classical action of gravity is replaced by the effective
action $\Gamma_{\rm eff} \left[g_{\mu\nu}\right]$, 
that is a functional of quantum matter fields $\phi$ in the 
classical background metric,
\footnote{
In theory involving gravitational fields, 
anomalies may exist and break general covariance or local Lorentz invariance. 
These are called gravitational anomalies and generally exist
when the spacetime dimension is $D = 4k + 2$, $k =0, 1, 2\dots$~\cite{AlvarezGaume:1983ig}. 
For four-dimensional semiclassical gravity they do not appear in general. 
}
\begin{align}\label{eq:gravity}
e^{i\Gamma_{\rm eff} \left[g_{\mu\nu}\right]}\,
=e^{iS_{\rm G}\left[g_{\mu\nu}\right]}\int { \mathcal{D}\phi \,e^{iS_{\rm M}\left[\phi,\,g_{\mu\nu}\right]} } ,
\end{align}
where $S_{\rm G}\left[g_{\mu\nu}\right]$ is the gravitational action
and $S_{\rm M}\left[g_{\mu\nu}\right]$ is the classical action of matter.
This procedure also corresponds to the large $N$ approximation
of quantum gravity~\cite{Hartle:1981zt}.
When quantum corrections of graviton is smaller than the
corrections of large number of matter fields, 
one can neglect the graviton loops. In particular, if one considers the early universe, 
there were certainly a large number of matter fields.
Thus, the large $N$ approximation is applicable 
and semiclassical gravity is valid.
For the current Universe, 
the corresponding energy scale is much smaller than the Planck scale
and the semiclassical approximation should be valid.

The gravitational action is constructed by 
the Einstein-Hilbert term and the cosmological constant,
\begin{equation}
S_{\mathrm{EH}}\left[g_{\mu\nu}\right]\equiv  
-\frac { 1 }{ 16\pi { G }_{ N } } \int { d }^{ 4 }x\sqrt { -g } \left( R+2\Lambda\right) \, ,
\end{equation}
and the fourth-derivative curvature terms,
\begin{align}
\begin{split}
S_{\mathrm{HD}}\left[g_{\mu\nu}\right]&\equiv \int{ d }^{ 4 }x\sqrt { -g }\,
\Bigl( { c }_{ 1 }R^2+{ c }_{ 2 }R_{\mu\nu}R^{\mu\nu} \\
&+{ c }_{ 3 }R_{\mu\nu\kappa\lambda} R^{\mu\nu\kappa\lambda}
+{ c }_{ 4 }\Box R \Bigr)  ,
\end{split}
\end{align}
The fourth-derivative terms
are indispensable for renormalization 
to eliminate one-loop divergences in curved spacetime.
Without the higher derivative terms
the semiclassical gravity becomes non-renormalizable
and is not consistent as fundamental (not effective) quantum field theory.
If one regard the semiclassical gravity as the fundamental,
higher derivative curvatures always exist at the classical level.
Even if these terms are not included into the classical action,
they will emerge from quantum energy momentum tensors. 
The effective action of Eq.~(\ref{eq:gravity}) 
derives the semiclassical Einstein's equations~\cite{Birrell:1982ix},
\begin{align}\label{eq:classical}\begin{split}
\frac { 1 }{ 8\pi { G }_{ N } }& \left( 
{ R }_{ \mu \nu  }-\frac { 1 }{ 2 }R{ g }_{ \mu \nu  }
+{ \Lambda  } { g }_{ \mu\nu }\right) \\ &\quad
+{ a }_{ 1 }{ H }_{ \mu\nu }^{ \left( 1 \right)  }
+{ a }_{ 2 }{ H }_{ \mu\nu  }^{ \left( 2 \right)  }
+{ a }_{ 3 }{ H }_{ \mu\nu  }=\left< { T }_{ \mu \nu  } \right> ,
\end{split}
\end{align}
where $\left< { T }_{ \mu \nu  } \right>$ is the vacuum expectation value of 
the quantum energy momentum tensor, 
\begin{align}\label{eq:tensor}\begin{split}
\left< { T }_{ \mu \nu  } \right>=-\frac{2}{\sqrt{-g}}\frac{\delta
\Gamma_{\rm eff} \left[g_{\mu\nu}\right]}{\delta g^{ \mu \nu  }} ,
\end{split}
\end{align}
and the geometric tensors $H^{(1,2)}_{\mu\nu}$ are defined as 
\begin{align}
\begin{split}
H^{(1)}_{\mu\nu} &\equiv 
\frac { 1 }{ \sqrt { -g }  } \frac { \delta  }{ \delta { g }^{ \mu \nu  } } \int { d } ^{ 4 }x\sqrt { -g } { R }^{ 2 } \\ &=
2\nabla_\nu \nabla_\mu R -2g_{\mu\nu}\Box R
 - {1\over 2}g_{\mu\nu} R^2 +2R R_{\mu\nu},   \\
H^{(2)}_{\mu\nu} &\equiv
\frac { 1 }{ \sqrt { -g }  } \frac { \delta  }{ \delta { g }^{ \mu \nu  } } \int { d } ^{ 4 }x\sqrt { -g }
R_{\mu\nu}R^{\mu\nu}\\ &=
2\nabla_\alpha \nabla_\nu R_\mu^\alpha - \Box R_{\mu\nu} -{1\over 2}g_{\mu\nu}\Box R
\\ &\quad\quad  -{1\over 2}g_{\mu\nu} R_{\alpha\beta}R^{\alpha\beta} 
+2R_\mu^\rho R_{\rho\nu},   \\
H_{\mu\nu} &\equiv 
\frac { 1 }{ \sqrt { -g }  } \frac { \delta  }{ \delta { g }^{ \mu \nu  } } \int { d } ^{ 4 }x\sqrt { -g }
R_{\mu\nu\kappa\lambda}R^{\mu\nu\kappa\lambda}\\
&= -  H^{(1)}_{\mu\nu} + 4H^{(2)}_{\mu\nu}.
\end{split}
\end{align}
For the flat FLRW universe, the geometrical tensors $H^{(1)}_{\mu\nu}$ and
$H^{(2)}_{\mu\nu}$ has a relation $H^{(1)}_{\mu\nu}=3H^{(2)}_{\mu\nu}$,
and we get,
\begin{align}
\begin{split}
&\quad\quad { a }_{ 1 }{ H }_{ \mu\nu }^{ \left( 1 \right)  }
+{ a }_{ 2 }{ H }_{ \mu\nu  }^{ \left( 2 \right)  }+{ a }_{ 3 }{ H }_{ \mu\nu  } \\
&=\left({ a }_{ 1 }+\frac{1}{3}{ a }_{ 2 }
+\frac{1}{3}{ a }_{ 3 }\right){ H }_{ \mu\nu }^{ \left( 1 \right)  }
={ \alpha }_{ 1 }{ H }_{ \mu\nu }^{ \left( 1 \right)  },
\end{split}
\end{align}
where we note ${ \alpha }_{ 1 }={ a }_{ 1 }
+\frac{1}{3}{ a }_{ 2 }+\frac{1}{3}{ a }_{ 3 }$.
Furthermore, the quantum energy momentum tensor $\left< { T }_{ \mu \nu  } \right>$
introduce more additional geometric tensors (for the detailed discussion
see  Ref~\cite{Birrell:1982ix}).
For instance, the renormalized vacuum energy momentum tensor
for a massless conformally-coupled scalar field is given by the
conformal anomaly,
\begin{align}
\left< { T }_{ \mu \nu  } \right>_{\rm conformal}=\frac{1}{2880\pi^2} 
\left(-\frac{1}{6}{ H }_{\mu \nu   }^{ \left( 1 \right)}
+ { H }_{ \mu \nu   }^{ \left( 3 \right)} \right) 
\end{align}
where 
\begin{align*}
H^{(3)}_{\mu\nu} 
&\equiv  \frac{1}{12}R^2g_{\mu\nu}-R^{\rho\sigma}R_{\rho\mu\sigma\nu}  \\
&=R_\mu^\rho R_{\rho\nu} - \frac{2}{3}R R_{\mu\nu}
  -\frac{1}{2}R_{\rho\sigma}R^{\rho\sigma}g_{\mu\nu}  
  +\frac{1}{4}R^2g_{\mu\nu} ,     
\end{align*}
The semiclassical gravity introduces 
more additional geometric tensor which depends on quantum states~\cite{Birrell:1982ix}.
In fact, the renormalized energy momentum tensor for a massless 
minimally-coupled scalar field
in Bunch-Davies vacuum state is given by~\cite{Bunch:1978yw},
\begin{align}\label{eq:bd}
\begin{split}
&\left< { T }_{ \mu\nu }\right>_{\rm ren}= \frac{\left(-\frac{1}{6}{ H }_{\mu \nu   }^{ \left( 1 \right)}
+ { H }_{ \mu \nu   }^{ \left( 3 \right)} \right) }{2880\pi^2} 
-\,\frac{{ H }_{ \mu \nu }^{ \left( 1 \right)}\log \left( \frac{R}{\mu^2} \right)}{1152\pi^2}  \\
& +\frac{\left(-32{ \nabla  }_{ \nu  }{ \nabla  }_{ \mu  }R
+56\Box Rg_{\mu\nu} - 8R R_{\mu\nu}+11R^2g_{\mu\nu} \right)}{13824\pi^2} .
\end{split}
\end{align}
which have higher derivative corrections.

In the next subsection we review the renormalization of 
the quantum energy momentum tensor
and see how the higher derivative curvatures  
appear in semiclassical gravity.

\subsection{Quantum Back-reaction}
\label{sec:backreaction}
First, we consider adiabatic (WKB) approximation for the conformally massless fields
and derive the renormalized quantum energy momentum tensors in this method.
It is found that the derivation of adiabatic (WKB) approximation
reduces the ambiguity of the UV divergences in renormalization and 
it is more significant than any other regularization
in curved spacetime.

In this paper, we consider a spatially flat 
Friedmann-Lemaitre-Robertson-Walker (FLRW) spacetime,
\begin{align}\label{eq:FLRW}
ds^{2}=dt^{2}-a^{2}\left(t\right)\delta_{ij}dx^{i}dx^{j},
\end{align}
where $a\left(t\right)$ is the scale factor and $t$ is the cosmic time.
We introduce conformal time $\eta$ defined by $d\eta=dt/a$.

Let us consider the matter action for the conformally coupled scalar field $\phi$
with mass $m$,
\begin{align}
S_{\rm M}=\int { { d }^{ 4 }x\sqrt { -g } \left( -\frac { 1 }{ 2 } { g }^{ \mu \nu  }
{ \partial   }_{ \mu  }\phi {\partial  }_{ \nu  }\phi -\frac{1}{2}\left(m^{2}+\frac{R}{6} \right)\phi^{2}  \right)  } ,
\end{align}
which leads to the Klein-Gordon equation given as 
\begin{align}\label{eq:Klein-Gordon}
\Box \phi -\left(m^{2}+ \frac{R}{6} \right)\phi  =0 .
\end{align}

The operator $\phi\left(\eta ,x\right)$ can be decomposed as
\begin{align}
\phi\left(\eta ,x\right) =\int { \frac{{ d }^{ 3 }k}{ { \left( 2\pi  \right)  }^{ 3/2 }}
\left( { a }_{ k }\frac { { e }^{ ik\cdot x }\varphi_{ k }\left( \eta  \right) }{a\left(\eta\right)  } 
+{ a }_{ k }^{ \dagger  }
\frac { { e }^{ -ik\cdot x }\varphi^{ * }_k\left(\eta  \right) }{a\left(\eta \right)  }   \right)  }  \label{eq:ddfkkfledg},
\end{align}
where ${ a }_{ k }$, ${ a }^{ \dagger  }_{ k }$ are the annihilation and creation 
operators respectively. 
In curved spacetime quantum states are determined by the choice of the mode functions. 
The mode function $\varphi_{ k }\left( \eta \right) $ 
should satisfy the Wronskian condition,
\begin{align}
\varphi_{ k }'^{ * }\left( \eta \right) \varphi_{ k }\left( \eta  \right)
-\varphi_{ k }'\left( \eta  \right)\varphi_{ k }^{ * }\left( \eta  \right)
= i\label{eq:dddrrrg},
\end{align}
which ensures the canonical commutation relations.
We adopt adiabatic (WKB) approximation to the mode function $\phi(\eta, x)$
which is written by~\cite{Parker:1974qw}:
\begin{align}
\begin{split}
\varphi_{ k }\left( \eta  \right) &=\frac { 1 }{ a\left( \eta  \right)\sqrt { 2{ W }_{ k }\left( \eta  \right)   }  } 
\Bigl( \alpha_{ k }\cdot e^{-i\int { { W }_{ k }\left( \eta  \right) \, d\eta  }} \\  &\quad +
\beta_{ k }\cdot e^{i\int { { W }_{ k }\left( \eta  \right)\, d\eta  }}\Bigr) \label{eq:jgskedg},
\end{split}
\end{align}
where the background changes slowly and must satisfy
the adiabatic (WKB) conditions (${ \omega  }_{ k }^{ 2 }>0$ and 
$\left| { \omega ' }_{ k }/{ \omega  }_{ k }^{ 2 } \right| \ll 1$ where 
${ \omega  }_{ k }^{ 2 }\left( \eta  \right) ={ k }^{ 2 }+a^2\left( \eta  \right){ m }^{ 2 } $).
The coefficients $\alpha_{ k }$ and $\beta_{ k }$ satisfy the following conditions, 
\begin{align}
{ { \left| { \alpha  }_{ k } \right|  }^{ 2 } }-{ \left| { \beta  }_{ k } \right|  }^{ 2 }=1.
\end{align}
The adiabatic function ${ W }_{ k }\left( \eta  \right)$ is given by
\begin{align}
\begin{split}
&{ W }_{ k }\left( \eta  \right)\simeq{ \omega  }_{ k }
-\frac { { m }^{ 2 }C }{ 8{ \omega  }_{ k }^{ 3 } } \left( D'+{ D }^{ 2 } \right) +\frac { 5{ m }^{ 4 }{ C }^{ 2 }{ D }^{ 2 } }{ 32{ \omega  }^{ 5 } }  \\
&+\frac { { m }^{ 2 }C }{ 32{ \omega  }_{ k }^{ 5 } } \left( D'''+4D'D+3{ D' }^{ 2 }+6D'D^{ 2 }+D^{ 4 } \right)  \\
&-\frac { { m }^{ 4 }{ C }^{ 2 } }{ 128{ \omega  }_{ k }^{ 7 } } \left( 28D''D+19{ D' }^{ 2 }+122{ D' }^{ 2 }+47D^{ 4 } \right)  \\ 
&+\frac { { 221m }^{ 6 }{ C }^{ 3 } }{ 256{ \omega  }_{ k }^{ 9 } } \left( D'D^{ 2 }+D^{ 4 } \right) 
-\frac { 1105{ m }^{ 8 }{ C }^{ 4 }{ D }^{ 4 } }{ 2048{ \omega  }_{ k }^{ 11 } }+\cdots \label{eq:oekgkf},
\end{split}
\end{align}
where $C(\eta)=a^2(\eta)$ and $D=C'/C$.
The mode function $\varphi_{ k }\left( \eta  \right)$
with $\alpha_{ k }=1$ and $\beta_{ k }=0$
is a reasonable choice for a sufficiently slow and smooth background~\cite{Parker:1974qw},
\begin{align}
{ \varphi }_k\left( \eta  \right)=\frac { 1 }{ \sqrt { 2{ W }_{ k }\left( \eta  \right) C\left( \eta  \right)  }  } 
\cdot e^{-i\int { { W }_{ k }\left( \eta  \right)\, d\eta  }},
\end{align}
which defines the adiabatic vacuum state
$\left| { \Psi }_{\rm A}  \right>$ which is annihilated by all the operators ${ a }_{ k }$.
Here we assume that the adiabatic vacuum state $\left| { \Psi }_{\rm A}  \right>$
is adequate initial vacuum in flat FLRW spacetime.

Let us consider the quantum energy momentum tensor.
The classical energy momentum tensor is given by~\cite{Bunch:1980vc}
\begin{align}
{ T }_{ \mu \nu  }&=\frac { -2 }{ \sqrt { -g }  } \frac { \delta S  }{ \delta { g }^{ \mu \nu  } }=
\frac{2}{3}{ \partial  }_{ \mu  }\phi { \partial  }_{ \nu  }\phi -\frac{1}{6}{ g }_{ \mu \nu  }{ g }^{ \rho \sigma  }{ { \partial  }_{ \rho  }\phi \partial  }_{ \sigma  }\phi \\ &-\frac{1}{3} \phi\nabla_{ \mu  }\nabla_{ \nu  }\phi+\frac{1}{3} { g }_{ \mu \nu  }{ \phi \Box \phi 
-\frac{1}{6} { G }_{ \mu \nu  }{ \phi  }^{ 2 }+\frac { 1 }{ 2 } { m }^{ 2 }{ g }_{ \mu \nu  }{ \phi  }^{ 2 } },\nonumber
\end{align}
and the corresponding trace
\begin{align}
T_{\phantom{\mu}\mu}^{\mu}={ m }^{ 2 }{ \phi  }^{ 2 } ,
\end{align}
which is exactly zero when $m\rightarrow 0$.
Thus, the conformally massless scalar field 
has the vanishing trace of the stress tensor classically.
It is found that the conformal invariance is broken in quantum field theory.
The vacuum expectation values  of the energy momentum tensor $\left< { T }_{ \mu \nu  } \right>$
for $\varphi_{ k }\left( \eta  \right)$ are given by
\begin{align}
\begin{split}
\left< { T }_{ 00 } \right> &=\frac { 1 }{ 4{ \pi  }^{ 2 }C\left( \eta  \right)  } \int { dk{ k }^{ 2 } } \biggl[ { \left| \varphi_{ k }'\left( \eta  \right)\right|  }^{ 2 }
+{ \omega  }_{ k }^{ 2 }{ \left| \varphi_{ k }\left( \eta  \right) \right|  }^{ 2 } \biggr], \\
\left< T_{\phantom{\mu}\mu}^{\mu} \right> &=\frac { 1 }{ 2{ \pi  }^{ 2 }{ C }^2\left( \eta  \right)  } \int { dk{ k }^{ 2 } } 
 \biggl[ C  m^{2}{ \left| \varphi_{ k }\left( \eta  \right) \right|  }^{ 2 }  \biggr],
\end{split}
\end{align}
where $\left< { T }_{ 00 } \right>$ is the time component of the energy 
momentum tensor and $\left< T_{\phantom{\mu}\mu}^{\mu} \right> $ is the trace.
These quantum energy momentum tensor has UV divergences and one must 
proceed the renormalization.

Next, we rewrite the vacuum expectation values $\left< { T }_{ \mu \nu  } \right>$ 
of the energy momentum tensor by the adiabatic approximation~\cite{Bunch:1980vc}
and renormalize the quantum energy momentum tensors as follows,
\begin{widetext}
\begin{align}
\begin{split}
\left< { T }_{ 00 } \right>&=\frac { 1 }{ 8{ \pi  }^{ 2 }C\left( \eta  \right)  } \int { dk{ k }^{ 2 } } \biggl[ 
2{ \omega  }_{ k }+\frac{C^{2}m^{4}D^{2}}{16{ \omega  }_{ k }^{ 5}}
-\frac{C^{2}m^{4}}{64{ \omega  }_{ k }^{ 7}}\left(2D''D-D'^{2}+4D'D^{2}+D^{4}\right) \\ &  
+\frac{7C^{3}m^{6}}{64{ \omega  }_{ k }^{ 9}}\left(D'D^{2}+D^{4}\right)-\frac{105C^{4}m^{8}D^{4}}{1024{ \omega  }_{ k }^{11}}  \biggr], \\  
\left< T_{\phantom{\mu}\mu}^{\mu} \right> &=\frac { 1 }{ 4{ \pi  }^{ 2 }C^{2}\left( \eta  \right)  } \int { dk{ k }^{ 2 } } \biggl[ 
\frac{Cm^{2}}{{ \omega  }_{ k }}+\frac{C^{2}m^{4}}{8{ \omega  }_{ k }^{5}}\left(D'+D^{2}\right)
-\frac{5C^{3}m^{6}D^{2}}{32{ \omega  }_{ k }^{7}} 
-\frac{C^{2}m^{4}}{32{ \omega  }_{ k }^{7}}\left(D'''+4D''D+3D'^{2}+6D'D^{2}+D^{4}\right)  \\ &
+\frac{C^{3}m^{6}}{128{ \omega  }_{ k }^{9}}\left(28D''D+21D'^{2}+126D'D^{2}+49D^{4}\right) 
 -\frac{231C^{4}m^{8}}{256{ \omega  }_{ k }^{11}}\left(D'D^{2}+D^{4}\right) +\frac{1155C^{5}m^{10}D^{4}}{2048{ \omega  }_{ k }^{13}}\biggr],
\end{split}
\end{align}
\end{widetext}
where the lowest-order term of the quantum energy momentum tensor
$\left< { T }_{ \mu \nu  } \right>$ actually diverges,
\begin{align}
\left< { T }_{ 00 } \right>_{\rm diverge}=\frac { 1 }{ 4{ \pi  }^{ 2 }C\left( \eta  \right)  } \int { dk{ k }^{ 2 } }{ \omega  }_{ k }\rightarrow \infty.
\end{align}
By adopting the dimensional regularization,
the quantum energy momentum tensor $\left< { T }_{ \mu \nu  } \right>$
can be regularized,
\begin{align}
\begin{split}
&\left< { T }_{ 00 } \right>_{\rm reg} =-\frac { { m }^{ 4 }C}{ 64{ \pi  }^{ 2 } } \left[ \frac { 1 }{ \epsilon  } 
+\frac { 3 }{ 2 } -\gamma +\ln { 4\pi  } +\ln { \frac { { \mu  }^{ 2 } }{ { m }^{ 2 } }  }  \right] \\
&+\frac { { m }^{ 2 }{ D }^{ 2 } }{ 384{ \pi  }^{ 2 } }-\frac { 1 }{ 2880{ \pi  }^{ 2 }{ C }} 
\left( \frac { 3 }{ 2 } D''D-\frac { 3 }{ 4 } { D' }^{ 2 }-\frac { 3 }{ 8 } { D }^{ 4 } \right),\\
&\left< T_{\phantom{\mu}\mu}^{\mu}\right>_{\rm reg} 
=-\frac { { m }^{ 4 }}{ 32{ \pi  }^{ 2 }C } \left[ \frac { 1 }{ \epsilon  } 
+1 -\gamma +\ln { 4\pi  } +\ln { \frac { { \mu  }^{ 2 } }{ { m }^{ 2 } }  }  \right] \\
&+\frac { { m }^{ 2 }{ D }^{ 2 } }{ 192{ \pi  }^{ 2 }C }\left(2D'+D^2\right) -\frac { 1 }{ 960{ \pi  }^{ 2 }{ C }^{ 2 } } \left( D'''-D'{ D }^{ 2 } \right) .
\end{split}
\end{align}
where $\mu$ is the renormalization parameter and 
$\gamma$ is the Euler-Mascheroni constant.
The $1/\epsilon$ terms represent the UV divergences and  
they must be absorbed by the counter-terms of the gravitational action.

Hence, the renormalized energy momentum tensor for flat spacetime is
\begin{align}\label{eq:trace-anomaly}
\begin{split}
&\left< { T }_{ 00 } \right>_{\rm ren}
= \frac { { m }^{ 4 }C}{ 64{ \pi  }^{ 2 } }
\left(\ln { \frac { { m  }^{ 2 } }{ { \mu }^{ 2 } }  }-\frac{3}{2}\right) 
+\frac { { m }^{ 2 }{ D }^{ 2 } }{ 384{ \pi  }^{ 2 }}\\
&\quad -\frac { 1 }{ 2880{ \pi  }^{ 2 }{ C }} 
\left( \frac { 3 }{ 2 } D''D-\frac { 3 }{ 4 } { D' }^{ 2 }-\frac { 3 }{ 8 } { D }^{ 4 } \right),\\
&\left< T_{\phantom{\mu}\mu}^{\mu}\right>_{\rm ren} 
=\frac { { m }^{ 4 }}{ 32{ \pi  }^{ 2 }C } \left( \ln { \frac { { m  }^{ 2 } }{ { \mu }^{ 2 } }  } -1 \right)\\
&+\frac { { m }^{ 2 }{ D }^{ 2 } }{ 192{ \pi  }^{ 2 }C }\left(2D'+D^2\right) -\frac { 1 }{ 960{ \pi  }^{ 2 }{ C }^{ 2 } } \left( D'''-D'{ D }^{ 2 } \right) .
\end{split}
\end{align}
where the first terms are the running 
cosmological constant corrections which
originates from the lowest adiabatic term.
On the other hand, the latter parts express
vacuum polarization or quantum particle creation in curved spacetime.
The anomaly term of Eq.~(\ref{eq:trace-anomaly}) is consistent with 
using dimensional regularization~\cite{Deser:1976yx,Duff:1977ay} and it 
is equal to $a_2(x)/16 \pi^2$~\cite{Buchbinder:1992rb}
where $a_2(x)$ is a coefficient of the DeWitt-Schwinger formalism. 
The conformal anomaly is given by the massless limit of Eq.~(\ref{eq:trace-anomaly}),
\begin{align}\label{eq:conformal-anomaly}
\begin{split}
\left< T_{\phantom{\mu}\mu}^{\mu}\right>_{\rm anomaly} &=\lim _{ m\rightarrow 0 }{ \left< T_{\phantom{\mu}\mu}^{\mu}\right>_{\rm ren}  } \\ &
=-\frac { 1 }{ 960{ \pi  }^{ 2 }{ C }^{ 2 } } \left( D'''-D'{ D }^{ 2 } \right) \\ &
=-\frac { 1 }{ 2880{ \pi  }^{ 2 } } \left[ \left( { R }_{ \mu \nu  }{ R }^{ \mu \nu  }-\frac { 1 }{ 3 } { R }^{ 2 } \right) +\Box R \right] \\ &
=\frac{1}{360(4\pi)^{2}}E - \frac{1}{180(4\pi)^{2}}\Box R ,
\end{split}
\end{align}
By using the adiabatic approximation we obtain the following expression
for a massless fermion~\cite{Landete:2013lpa,delRio:2014cha},
\begin{align}\label{eq:conformal-anomaly-f}
\begin{split}
\left< T_{\phantom{\mu}\mu}^{\mu}\right>_{\rm anomaly}^{\rm fermion}
&=-\frac { 1 }{ 2880{ \pi  }^{ 2 } } \left[ 11\left( { R }_{ \mu \nu  }{ R }^{ \mu \nu  }-\frac { 1 }{ 3 } { R }^{ 2 } \right) + 6\, \Box R \right] \\ &
=\frac{11}{360(4\pi)^{2}}E - \frac{6}{180(4\pi)^{2}}\Box R.
\end{split}
\end{align}
The conformal anomaly for the gauge field in adiabatic expansion
is given by~\cite{Chu:2016kwv}:
\begin{align}\label{eq:conformal-anomaly-g}
\begin{split}
&\left< T_{\phantom{\mu}\mu}^{\mu}\right>_{\rm anomaly}^{\rm gauge\, boson}
=-\frac { 1 }{ 2880{ \pi  }^{ 2 } } \biggl[ 62\left( { R }_{ \mu \nu  }{ R }^{ \mu \nu  }
-\frac { 1 }{ 3 } { R }^{ 2 } \right) 
\\ &- \left(18+15 \log \xi \right) \Box R \biggr] 
=\frac{62}{360(4\pi)^{2}}E + \frac{\left(18+15 \log \xi \right)}{180(4\pi)^{2}}\Box R.  
\end{split}
\end{align}
where $\xi$ is a gauge fixing parameter defined by the covariant gauge fixing term~\cite{Chu:2016kwv}:
\begin{align}
\mathcal{ L  }_{ \rm gf }=-\frac { \sqrt { -g }  }{ 2\xi  } { \left( { \nabla  }^{ \mu  }{ A }_{ \mu  } \right)  }^{ 2 }.
\end{align}
The gauge dependence of Eq.~(\ref{eq:conformal-anomaly-g}) 
also exists in the DeWitt-Schwinger expansion
formalism~\cite{Endo:1984sz,Toms:2014tia,Vieira:2015oka}. The adiabatic approximation reproduces the 
gauge dependence of the $\Box R$ term which has also the regularization-scheme dependence.
However, the gauge fixing parameter can be removed by the gravitational coupling constants
in the Einstein equation and we can drop the gauge fixing parameter $\xi$.
It is found out that the adiabatic expressions for the conformal anomaly precisely matches the expression derived by effective action using the dimensional regularization~\cite{Buchbinder:1992rb}.

The renormalized energy momentum tensor $\left< { T }_{ \mu\nu } \right>_{\rm ren}$
are generally given as follows,
\begin{align}\label{eq:semiclassical-tensor}
\begin{split}
\left< { T }_{ \mu\nu }\right>_{\rm ren}&={ \alpha }_{ 1 }{ H }_{ \mu\nu }^{ \left( 1 \right)  }
+{ \alpha }_{ 3 }{ H }_{ \mu\nu  }^{ \left( 3 \right)  }+{ \alpha }_{ 4 }{ H }_{ \mu\nu  }^{ \left( 4 \right)  }\end{split}
\end{align}
where the geometric tensor ${ H }_{ \mu \nu   }^{ \left( 4 \right)}$
depends on quantum states~\cite{Birrell:1982ix}
and the above equations are defined as fourth-order derivative equations.
The dimensionless parameters ${ \alpha }_{ 1,3 }$ 
for massless fields are given by~\cite{Birrell:1982ix}, 
\begin{subequations}
\label{eq:alpha}
\begin{align}
{ \alpha }_{ 1 } &= \frac{-1}{2880\pi^{2}}\left( \frac{N_{S}}{6}+N_{F}-3N_{G}\right), \\
{ \alpha }_{ 3} &= \frac{1}{2880\pi^{2}}\left( N_{S}+\frac{11}{2}N_{F}+ 62 N_{G}\right), 
\end{align}
\end{subequations}
where we consider $N_{S}$ scalars (spin-0), $N_{F}$ Dirac fermions (spin-1/2) 
and $N_{G}$ abelian gauge fields (spin-1). 
For instance, the Minimal Supersymmetric Standard Model (MSSM), 
takes the following values: $N_{S}=104$, $N_{F}=32$ and $N_{G}=12$.
On the other hand, the Standard Model (SM) takes the following values: $N_{S}=4$,
$N_{F}=24$ and $N_{G}=12$ where the right-handed neutrinos are assumed.
Finally, the current universe has only photon, 
$N_{S}=0$, $N_{F}=0$ and $N_{G}=1$.
It is found that the large number of the 
scalar fields or fermions lead to the negative ${ \alpha }_{ 1}$ 
and induce the runaway solutions.

\subsection{Covariant Conservation Laws}
\label{sec:backreaction}
Briefly, we comment covariant conservation laws for 
the quantum energy momentum tensor~\cite{Davies:1977ti}.
The energy momentum tensor in both classical and quantum cases must satisfy the 
covariant conservation laws,
\begin{align}\label{eq:conservation}
\nabla_{ \mu } \left< { T }^{ \mu\nu }\right>_{\rm ren}=0,
\end{align}
which means energy and momentum conservations.
We rewrite the trace of Eq.~(\ref{eq:semiclassical-tensor}) 
for massless conformally-invariant fields in terms of the scale factor $a(t)$,
\begin{align}\label{eq:conf-anomaly}
\left< T_{\phantom{\mu}\mu}^{\mu}\right>_{\rm ren}&=-36\alpha{ a }^{ -3 }\left[ { a }^{ 2 }{ a }^{ (4) }+3a\dot { a } { a }^{ (3) }+a\ddot { a } ^{ 2 }-5{ \dot { a }  }^{ 2 }\ddot { a }  \right]\nonumber \\
 &+12\beta { a }^{ -3 }\ddot { a } { \dot { a }  }^{ 2 },
\end{align}
where the dots and bracketed superscripts denote differentiation with respect to $t$.
Using the above expression, Eq.~(\ref{eq:conservation})
derives the renormalized vacuum energy density,
\begin{align}\label{eq:conf-anomaly}
\rho_{\rm ren}&=-36\alpha{ a }^{ -4 }\left[ { a }^{ 2 }\dot { a }{ a }^{ (3) }+a\dot{ a }^{2}\ddot{ a }
-\frac{1}{2}{ a }^{ 2 }\ddot{ a } ^{ 2 }-\frac{3}{2}\dot{ a }^{ 4 } \right]\nonumber \\
 &+3\beta { a }^{ -4 }{ \dot { a }  }^{ 4 }+C{ a }^{ -4 },
\end{align}
where $C$ is a constant from integration and the last term 
corresponds to the thermal radiation $\rho\propto { a }^{ -4 }$.
Hence, the renormalized expression of 
the energy momentum tensor of Eq.~(\ref{eq:semiclassical-tensor}) 
satisfies the covariant conservation laws.
From here, we drop the constant $C$ for simplicity.

\subsection{Gravitational Thermodynamics and Averaged Null Energy Condition}
\label{sec:thermodynamics}
In this section let us briefly discuss several problems of the semiclassical gravity
such as violations of the gravitational thermodynamical law 
which is characterized by thermodynamical entropy $S$ and  
the averaged null energy condition. 
Famously, the black hole thermodynamics assumes that black holes have the entropy,
quantified by the area of the event horizon,
\begin{equation}
S_{\rm BH} = \frac{A}{4G_{N}},
\end{equation}
where the horizon area $A$ is quantified by the surface gravity 
$\kappa$ and the mass $M_{\rm BH}$ of stationary black holes,
\begin{equation}
dM_{\rm BH} = \frac{\kappa dA}{8\pi G_{N}}+
(\textrm{rotation \& charge terms}).
\end{equation}
Classically, the black holes acquire 
mass from other massive objects and $S_{\rm BH}$ always increases.
This fact matches the thermodynamical interpretation of the entropy.
However, the black holes may lose its mass due to the Hawking radiation
with the temperature,
\begin{equation}
T_{\rm H} = \frac{\kappa}{2\pi},
\end{equation}
and thus $S_{\rm BH}$ decreases. 
On the other hand, thermal character of the event horizon in de Sitter space
formally defines de Sitter entropy,
\begin{equation}
{S_{\rm dS}} = \frac{\pi H^{-2}}{G_{N}},
\end{equation}
where the horizon area is given by $A=4\pi H^{-2}$
and the time-evolution is written as follows,
\begin{equation}
\frac{dS_{\rm dS}}{dt} = -\frac{2\pi H^{-3}\dot{H}}{G_{N}}.
\end{equation}

By using the de Sitter entropy $S_{\rm dS}$ one can get interesting consequences
such as a no-go theorem for slow-roll eternal inflation. 
Let us consider slow-roll inflation driven by a inflaton field $\phi$. 
For the slow-roll inflation, we have
\begin{equation}\label{eq:Hubb}
\dot { H } =-4\pi { G }_{ N }\dot{\phi}^2 \,.
\end{equation}
Hence, the de Sitter entropy $S_{\rm dS}$ is rewritten as,
\begin{equation}
\frac{dS_{\rm dS}}{dN} = -\frac{2\pi \dot{H}}{G_{N}H^{4}}
=\frac{8\pi^2 \dot{\phi}^2}{H^{4}}\sim \left(\frac{\delta\rho}{\rho}\right)^{-2}
\gtrsim 1,
\end{equation}
where $N$ is the number of e-foldings defined by  $dN = Hdt$, 
$\rho$ is the energy density and $\delta\rho$ is the
energy density perturbation satisfying $|\delta\rho/\rho|\lesssim 1$.
The total number of e-folding $N_{\rm tot}$ is bounded as follows~\cite{ArkaniHamed:2007ky}, 
\begin{equation}
N_{\rm tot}\lesssim  \Delta S =S_{\rm end}-S_{\rm ini},
\end{equation}
where $S_{\rm end}$ and $S_{\rm ini}$ are the de Sitter entropy 
at the end and the beginning of the inflation, respectively. 
For the large field inflation, the entropy at the beginning 
is much smaller than that at the end, and then one get 
$\Delta S\sim S_{\rm end}$ but $\Delta S \ll S_{\rm end}$ 
for the small field inflation. 
In any case the total e-folding number $N_{\rm tot}$ is strictly restricted.

Let's discuss a more general case.
For flat FLRW universe,
the Friedmann equations yield a simple equation,
\begin{equation}\label{eq:Hubb}
\dot { H } =-4\pi { G }_{ N }\left( \rho+P  \right) \,.
\end{equation}
Hence, the de Sitter entropy can be written as follows,
\begin{equation}
\frac{dS_{\rm dS}}{dt} = 8\pi^2 H^{-3}\left( \rho+P  \right)\ge  0,
\end{equation}
which always increases and matches gravitational thermodynamical laws
when the null energy condition (NEC) is satisfied,
\begin{align}
T_{\mu\nu}k^{\mu}k^{\nu}\ge  0 \ \Longrightarrow \ \rho+P \ge  0,
\end{align}
where $k^{\mu}$ is the null (light-like) vector.
It is known that the NEC and the gravitational thermodynamical laws are 
closely related with each other~\cite{ArkaniHamed:2007ky}.
In general relativity, the NEC is a necessary condition to
eliminate any pathological spacetime or unphysical consequences such as wormhole,  
geometric instability and superluminal propagation. 
It is well known that the classical matters always satisfy the NEC and in this sense the 
classical general relativity does not violate any gravitational principles.
However, these classical conditions can be easily violated in quantum field theory (QFT)
and the NEC is broken even for quantum fields in Minkowski spacetime~\cite{Epstein:1965zza}
 (e.g. squeezed vacuum states~\cite{Kuo:1996jd}).
More generally, the averaged null energy condition (ANEC)~\cite{Borde:1987qr},
which is satisfied in Minkowski spacetime~\cite{Klinkhammer:1991ki} and prohibits 
a traversable wormhole~\cite{Friedman:1993ty}
have been proposed, 
 \begin{align}
\int _{ \gamma  }^{  }{ 
T_{\mu\nu}k^{\mu}k^{\nu}dl } 
\ge  0 \ \Longrightarrow \ 
\int _{-\infty   }^{ \infty  }\frac{1}{a}{\left(\rho+P\right)dt} \ge  0,
\end{align}
where the integral is taken over a null geodesic $\gamma$, 
$k^{\mu}$ is the parameterized tangent vector to the geodesic
and $l$ is the affine parameter~\cite{Li:2008sw}.
However, it has been known that curved spacetime, i.e.,
semiclassical gravity violates the NEC or ANEC~\cite{Visser:1994jb,
Visser:1996iv,Urban:2009yt,Urban:2010vr}
and the de Sitter entropy
decreases~\cite{Matsui:2019tah}.
This seems very plausible because the energy density undergoes 
quantum vacuum fluctuations and the variances of the energy density would be 
allowed to be both negative and positive in QFT.
Although so-called “quantum inequalities (QIs)~\cite{Ford:1978qya} has been proposed,
in principle, there is no lower limit for the negative vacuum energy 
and one can take any non-physical spacetime. 
In a nutshell, such quantum effects on gravity require careful discussion and
theory should not break these basic gravitational principles.

Finally, let us briefly see the violations of the NEC, ANEC and gravitational thermodynamical laws in
semiclassical gravity and simply consider conformal anomaly $\left< T_{\phantom{\mu}\mu}^{\mu}\right>_{\rm anomaly}$ in Eq.~(\ref{eq:Hubble-equation}). 
For the semiclassical gravity, the NEC, ANEC and de Sitter entropy 
can be violated with various conditions as follows;
\begin{align}
&\quad \rho+P  = 12\alpha_1
\left(6\dot { H }^{2 } +3H\ddot { H } + \dddot { H }\right)
-4 \alpha_3{ H }^{ 2 }\dot { H }\ngeq  0,\\
&\int _{-\infty   }^{ \infty  }\frac{1}{a}{\left(\rho+P\right)dt} = 
\int _{-\infty   }^{ \infty  }\biggl\{  \frac{12\alpha_1}{a}
\left(6\dot { H }^{2 } +3H\ddot { H } + \dddot { H }\right)\nonumber\\
&\quad\quad\quad\quad-\frac{4 \alpha_3}{a}{ H }^{ 2 }\dot { H }\biggr\}dt \ngeq  0,\\
&\frac{dS_{\rm dS}}{dt} = 
96\pi^2\alpha_1
\left(6{ H }^{-3}\dot { H }^{2 } +3H^{-2}\ddot { H } + H^{-3}\dddot { H }\right)\nonumber\\
&\quad\quad\quad\quad-32\pi^2 \alpha_3{ H }^{ -1 }\dot { H }\ngeq  0,
\end{align}
which suggests that semiclassical gravity is incompatible 
with basic gravitational principles~\cite{ArkaniHamed:2007ky}.

\section{Quantum spacetime instability}
\label{sec:instability}
We now turn to a more quantitative discussion of the validity of the semiclassical gravity.
The quantum energy momentum tensor holding 
higher derivative terms modifies the Einstein's equations
and destabilizes the classical solutions.
In this section, we investigate the quantum spacetime instabilities
in semiclassical gravity and consider the cosmological dynamics of the Universe.

First, we will revisit some results of Refs~\cite{Horowitz:1978fq,
Suen:1988uf,Suen:1989bg} and 
discuss the instability of the Minkowski spacetime 
under small perturbations.
Next, we discuss the instabilities of the homogenous and 
isotropic FLRW Universe, and consider the cosmological dynamics.
Our results suggest that the corresponding solutions 
of the semiclassical gravity can not be
incompatible with the cosmological observations.

For simplicity, we consider the semiclassical
Einstein’s equations for the massless conformally-invariant fields
and the classical radiation or non-relativistic matters,
\begin{align}\label{eq:semiclassical}
&\frac { 1 }{ 8\pi { G }_{ N } } \left( 
{ R }_{ \mu \nu  }-\frac { 1 }{ 2 }R{ g }_{ \mu \nu  }
+{ \Lambda  } { g }_{ \mu\nu }\right)=
\left< { T }_{ \mu\nu }\right>_{\rm ren}+{ T }_{ \mu\nu }^{\rm \,c}
\nonumber \\
&\quad ={ \alpha }_{ 1 }{ H }_{ \mu\nu }^{ \left( 1 \right)  }
+{ \alpha }_{ 3 }{ H }_{ \mu\nu  }^{ \left( 3 \right)  }+{ T }_{ \mu\nu }^{\rm \,c},
\end{align}
where ${ T }_{ \mu\nu }^{\rm \,c}$ is the energy momentum tensor
for ordinary radiation or non-relativistic matters.

\subsection{Instability of Minkowski spacetime 
under conformally flat perturbations }
Let us consider the Minkowski spacetime with no matter fields 
and investigate whether the spacetime is stable under the perturbations.
Although in classical general relativity, the Minkowski spacetime should be
completely stable, semiclassical gravity does not 
ensure this important fact. Refs~\cite{Horowitz:1978fq,Horowitz:1980fj,Hartle:1981zt,
RandjbarDaemi:1981wd,Jordan:1987wd,Suen:1988uf,Suen:1989bg,Anderson:2002fk} have
shown that the Minkowski spacetime is unstable under small perturbations
and the perturbations either grow exponentially or oscillate 
even in Planck time. This leads to disaster.
Before studying the FRLW instabilities and considering the influence on the 
current Universe, let us reconsider some results of the Minkowski instability 
of Ref~\cite{Horowitz:1978fq,Suen:1988uf,Suen:1989bg}.

In the Minkowski spacetime, the geometric tensors in semiclassical Einstein's equations satisfy,
\begin{align}\label{eq:Minkowski}
{ G }_{ \mu\nu  }={ H }_{ \mu\nu  }^{ \left( 1 \right)  }
={ H }_{ \mu\nu  }^{ \left( 3 \right)  }=0.
\end{align}
Now, we consider only conformally flat perturbations and write the metric as follows,
\begin{align}\label{eq:Minkowski}
{ g }_{ \mu\nu  }=\Omega^2{ \eta }_{ \mu\nu  },
\end{align}
where $\eta_{ \mu\nu  }$ is the Minkowski metric ($+---$ convention) and 
$\Omega$ is the conformal parameter.

The corresponding Ricci tensor $R_{ \mu\nu  }$ is given by 
\begin{align}\label{eq:Minkowski}
\begin{split}
R_{ \mu\nu  }&=4{ \Omega  }^{ -2 }\left( { \partial  }_{ \mu  }\Omega  \right) \left( { \partial  }_{ \nu  }\Omega  \right) -2{ \Omega  }^{ -1 }{ \partial  }_{ \mu  }{ \partial  }_{ \nu  }\Omega \\
&-{ \Omega  }^{ -1 }{ \eta  }_{ \mu \nu  }\left( { { \partial  } }^{ \alpha  }{ \partial  }_{ \alpha  }\Omega  \right) -{ \Omega  }^{ -2 }{ \eta  }_{ \mu \nu  }\left( { { \partial  } }^{ \alpha  }{ \Omega \partial  }_{ \alpha  }\Omega  \right) ,
\end{split}
\end{align}
where $\Omega=1$ reproduces the Minkowski spacetime solution.
Now, we rewrite Eq.~(\ref{eq:semiclassical}) using this expression with the
conformally flat perturbations $\Omega=1+\gamma$ and one can obtain~\cite{Horowitz:1978fq},
\begin{align}
\begin{split}
-&{ \partial  }_{ \mu  }{ \partial  }_{ \nu  }\gamma +\left( \Box \gamma  \right) { \eta  }_{ \mu \nu  }
+48\pi \alpha_{ 1 }G_N \\
&\times \left[- { \partial  }_{ \mu  }{ \partial  }_{ \nu  }\left( \Box \gamma  \right) +\Box \left( \Box \gamma  \right) { \eta  }_{ \mu \nu  } \right] =0,
\end{split}
\end{align}
where $\Box \equiv \eta^{\alpha \beta}{ \partial  }_{ \alpha  }{ \partial  }_{ \beta  }={ \partial  }^{ \alpha  }{ \partial  }_{ \alpha  }$.
This perturbation equation for $\gamma$ can be rewritten by a
simple equation,
\begin{align}
\left( { \partial  }_{ \mu  }{ \partial  }_{ \nu  }-{ \eta  }_{ \mu \nu  }\Box  \right) f=0,
\end{align}
where we define $f\equiv\left( \gamma +48\pi \alpha_{ 1 }G_N \Box \gamma  \right)$
and the general solution is clearly, 
\begin{align}
f={ k }^{ \alpha  }{ x }_{ \alpha  }+{\rm const.},
\end{align}
where ${ k }^{ \alpha  }$ expresses a constant vector field and 
${ x }^{ \alpha  }$ is a position vector field in the Minkowski spacetime.
Hence, we can get the following equation,
\begin{align}
\left(1+48\pi \alpha_{ 1 }G_N \Box \right)\gamma ={ k }^{ \alpha  }{ x }_{ \alpha  }+{\rm const.}
\end{align}
The most general solution to the above equation is 
\begin{align}
\gamma ={ k }^{ \alpha  }{ x }_{ \alpha  }+\chi+{\rm const.},
\end{align}
where inhomogeneous solution $\gamma ={ k }^{ \alpha  }{ x }_{ \alpha  }+{\rm const.}$
is pure gauge in the Minkowski spacetime
and the metic perturbation $\chi$ satisfies the Klein-Gordon equation,
\begin{align}
\Box\chi+\frac { 1 }{ 48\pi \alpha_{ 1 } G_N  }\chi=0,
\end{align}
which admits the spatially homogeneous solutions,
\begin{align}
\chi=C_1\sin\left(\omega t\right)+C_2\cos\left(\omega t\right), \ C_3 e^{t/\tau},
\end{align}
where $\omega=(48\pi \alpha_{ 1 }G_N)^{1/2}$ and 
$C_{1,2,3}$ are constants. The first solution is for $\alpha_{ 1 }>0$ and 
consistent with the ordinary Klein-Gordon equation.
However, if one takes $48\pi \alpha_{ 1 }\sim \mathcal{O}(1)$,
the perturbations oscillate in the Planck time
$t_{\rm P}=(G_N)^{1/2}=10^{-43}\ {\rm sec}$ 
and they emit the Planck energy photons, $E\sim 10^{19}\ {\rm GeV}$~\cite{Horowitz:1978fq}
which is unreasonable for the observed Universe.
The last possible solution is given 
for $\alpha_{ 1 }<0$. We denote $\tau=(48\pi \alpha_{ 1 } G_N)^{1/2}$.
This correspond to the Klein-Gordon equation with a negative mass
and suggests that the perturbations exponentially grow even in the Planck time.

Once the Minkowski spacetime is perturbed, 
the perturbations lead to a catastrophe.
For $\alpha_{ 1 }<0$ the instability time-scale can be summarized as,
\begin{align}
\begin{split}
t_{\rm I}&=(48\pi \alpha_{ 1 } G_N)^{1/2}=(48\pi \alpha_{ 1 })^{1/2}\cdot10^{-43}\ {\rm sec.}\\
&=(48\pi \alpha_{ 1 }/10^{118})^{1/2}\ {\rm Gyr.}
\end{split}
\end{align}
The instability time $t_{\rm I}$ must be as large as the age of the observed Universe,
$t_{\rm Age}=13.787 \pm 0.020\ {\rm Gyr.}$~\cite{Aghanim:2018eyx}
(Planck 2018, TT,TE,EE+lowE+lensing+BAO 68\% limits).
Otherwise, the perturbations or scale factor exponentially grow 
and our Universe is seriously destabilized.
Hence, we obtain the stable condition against quantum back-reaction,
\begin{align}\label{eq:condition}
\alpha_{ 1 }\gtrsim 10^{118},
\end{align}
which requires a large value of the gravitational curvature coupling or 
a large number of the particles species $\mathcal{N}\sim10^{118}$
for the high energy theory.
Hence, the $\alpha_{ 1 }<0$ case is trouble for the homogenous and isotropic
flat Universe. We will confirm these results 
in different methods in the next subsection
\footnote{ 
Taking the following conformal parameter,
\begin{align*}
{ g }_{ \mu\nu  }=\Omega^2{ \eta }_{ \mu\nu  }=a^2(\eta){ \eta }_{ \mu\nu  },
\end{align*}
one can get a similar consequence for the scale factor $a(\eta)$.}.

\subsection{Numerical analysis for Minkowski 
spacetime instability under perturbations}
Let us consider the FLRW spacetime with matter fields and 
study the spacetime insatiabilities from the quantum back-reaction
\footnote{
The similar analysis was given recently by one of the authors~\cite{Matsui:2019tah}.}.
For the Minkowski spacetime we have no classical matter and no 
cosmological constant.

By using Eq.~(\ref{eq:semiclassical})
we obtain the following semiclassical equation~\cite{Simon:1990jn},
\begin{align}\label{eq:R-equation}
\begin{split}
\frac{dR}{dt}=\frac{1}{12H}R^2-HR-\frac{H}{16\pi G_N\alpha_1}
+\frac{\alpha_3}{\alpha_1}H^3,
\end{split}
\end{align}
If we assume that the third term of Eq.~(\ref{eq:R-equation}) can dominate
due to the smallness of $16\pi G_N\alpha_1$, 
we can approximately rewrite,
\begin{align}\label{eq:Rt-equation}
\begin{split}
\frac{d^2H}{dt^2}\approx-\frac{1}{16\pi G_N\alpha_1}H,
\end{split}
\end{align}
where we consider $H(0)\approx 0$, $R(0)\approx 0$.
This admits the exponential or oscillating solutions for 
the Planck time $t_{\rm P}$
for $16\pi \alpha_{ 1 }\sim \mathcal{O}(1)$.
Thus, these are consistent with the previous discussion.
\begin{figure}[t]
\includegraphics[width=85
mm]{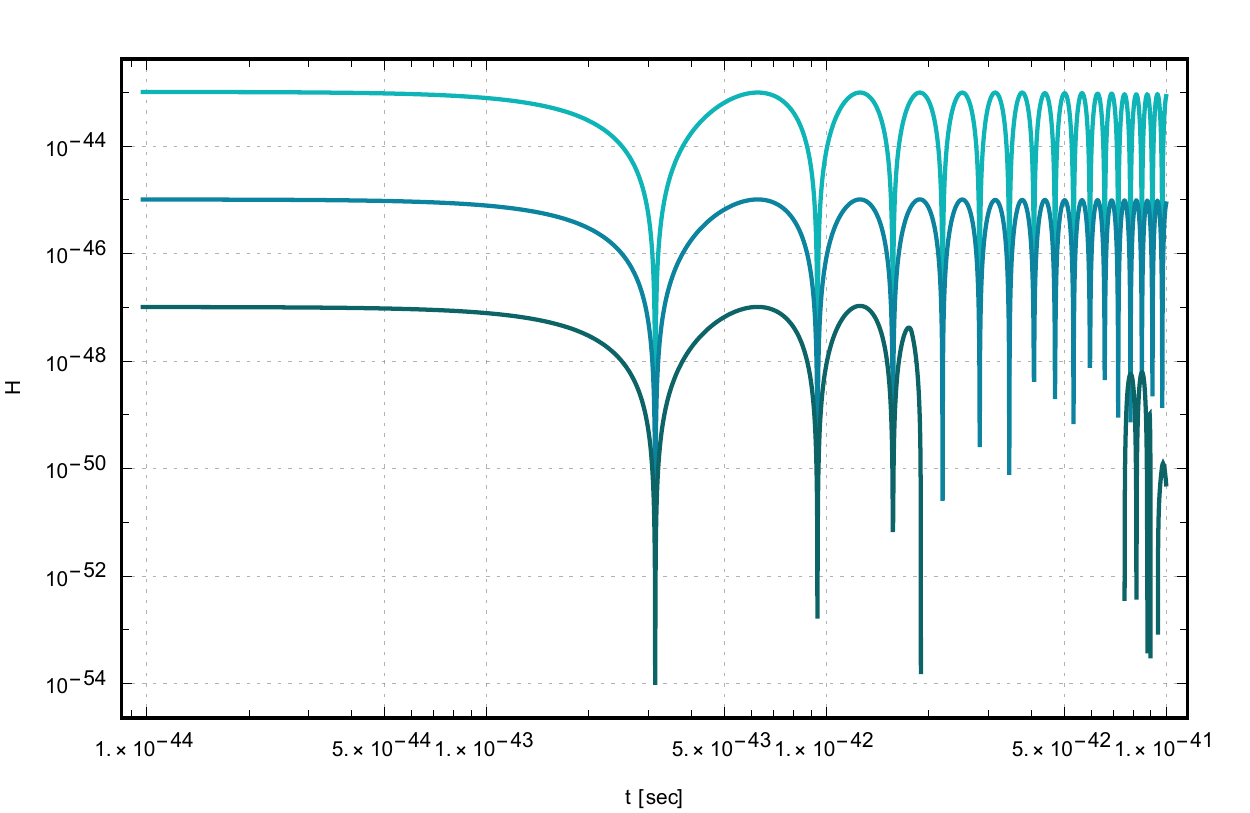}\\
\includegraphics[width=85
mm]{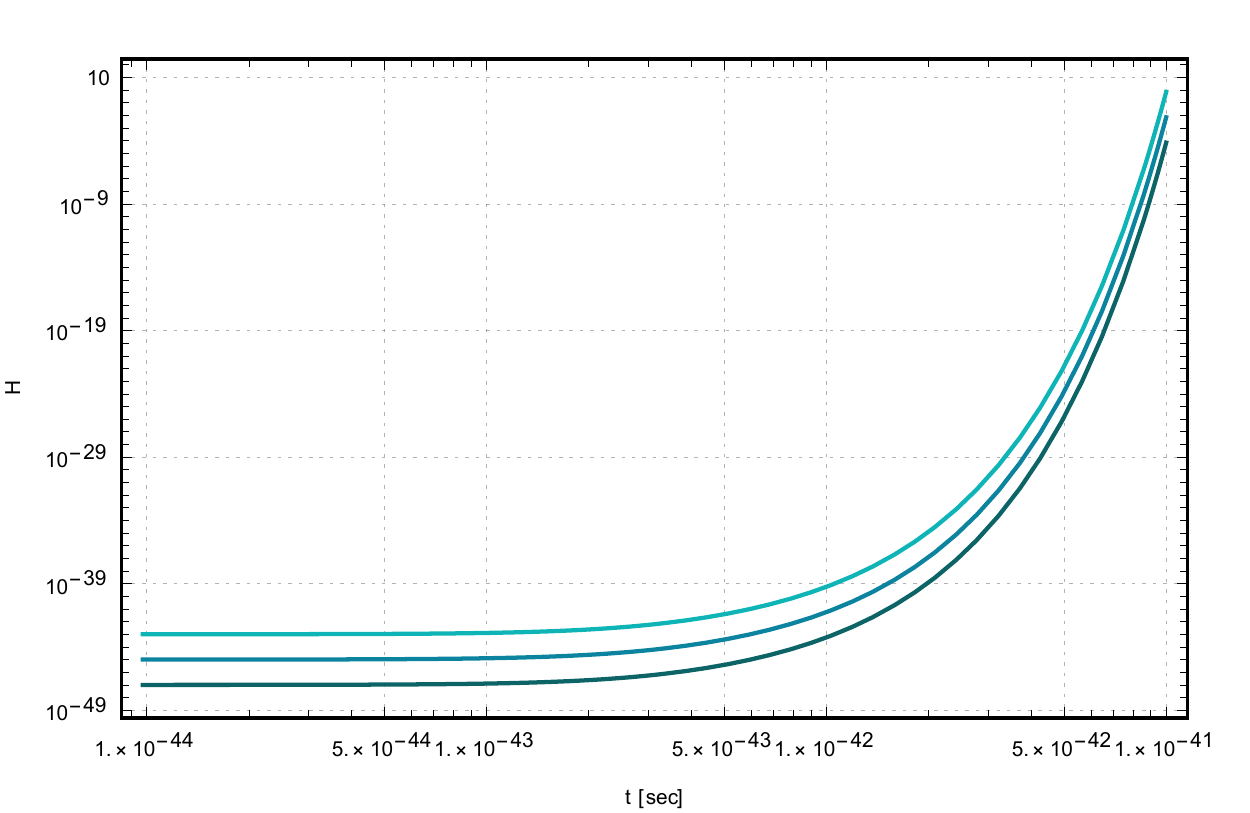}
\caption{We consider the instability of the Minkowski spacetime under perturbations
and show that the dynamics of the Hubble perturbation $H(t)$
around the Planck time $t_{\rm P}=10^{-43}\ {\rm sec}$.
We assume the initial conditions and the couplings of Eq.~(\ref{eq:Minkowski-initial}).
The top figure assumes $\alpha_{ 1 }=10^{-86}$ 
whereas the bottom figure takes $\alpha_{ 1 }=-10^{-86}$.
The top-to-bottom lines corresponds to $H_0=10^{-43,-45,-48}$,
respectively. }
\label{fig:Minkowski}
\end{figure}
Next, we numerically confirm the above analytical estimations. 
Let us assume the Minkowski spacetime perturbed by the small Hubble variations.
Using Eq.~(\ref{eq:R-equation}) and $R=6(\dot{H}+2H^2)$, 
we obtain the differential equation.
We investigate the system of equations starting at $t=0$ with
various conditions and perturbations. 
It is found that the numerical solutions of the system 
show that the Minkowski spacetime is unstable 
and they are consistent with the above analytical estimation.

In Fig.\ref{fig:Minkowski} 
we demonstrate numerical results for the Hubble perturbation $H(t)$ determined  by
Eq.~(\ref{eq:hubble}) with the following conditions and couplings,
\begin{align}
\begin{split}\label{eq:Minkowski-initial}
\textrm{Fig.\ref{fig:Minkowski}:}\ &H_0=10^{-43,-45,-48},\  \dot {H }_0=0,
\\ &  48\pi \alpha_{ 1 }G_N=\pm10^{-86},\ {8\pi G_N }\alpha_3=0, 10^{-86}, \\
\end{split}
\end{align}
where we set these parameters to respect the Planck time $t_{\rm P}=10^{-43}$
and it is found that magnitudes of the perturbation $\Upsilon(t)$
and values of $\alpha_{ 3 }$ are irreverent for the dynamics.
We found that the Hubble oscillations with the Planck frequency
occur for $\alpha_{ 1 }>0$ whereas for $\alpha_{ 1 }<0$,
the Hubble perturbations exponentially grow even in the Planck time $t_{\rm P}$.
The small values of $\alpha_{ 1 }$ leads to the faster destabilization and the 
stability of the Minkowski spacetime requires a large value of $\left|\alpha_{ 1 } \right|$.

\begin{figure*}[ht]
	\centering
	\includegraphics[scale=0.65]{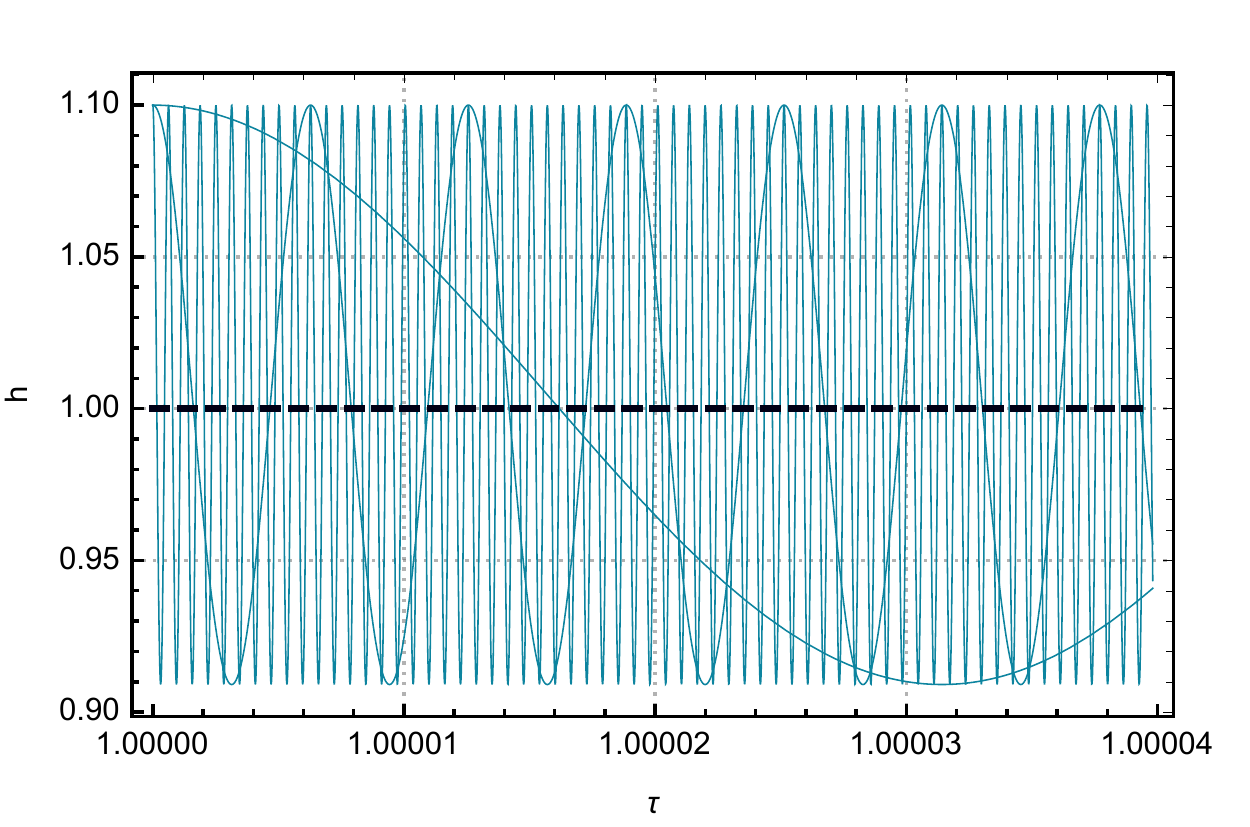}
	\includegraphics[scale=0.65]{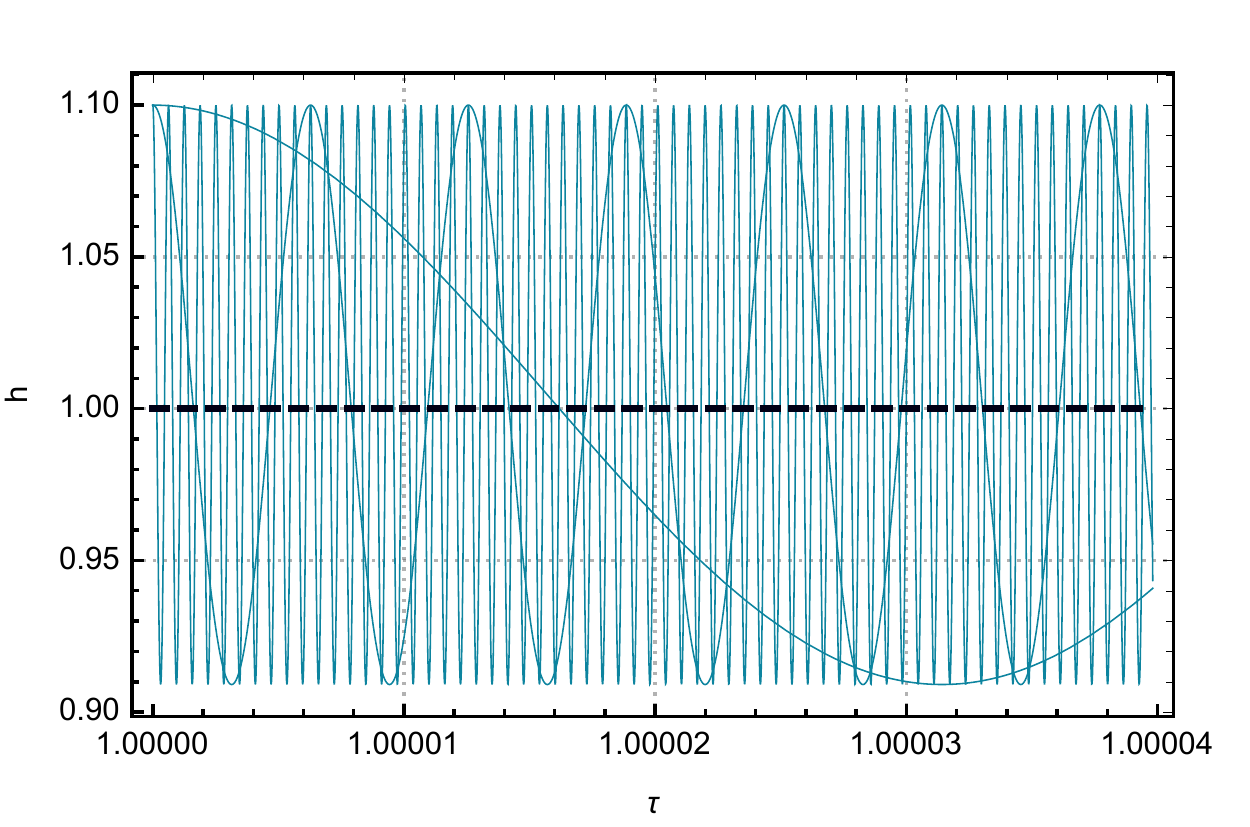}
	\includegraphics[scale=0.65]{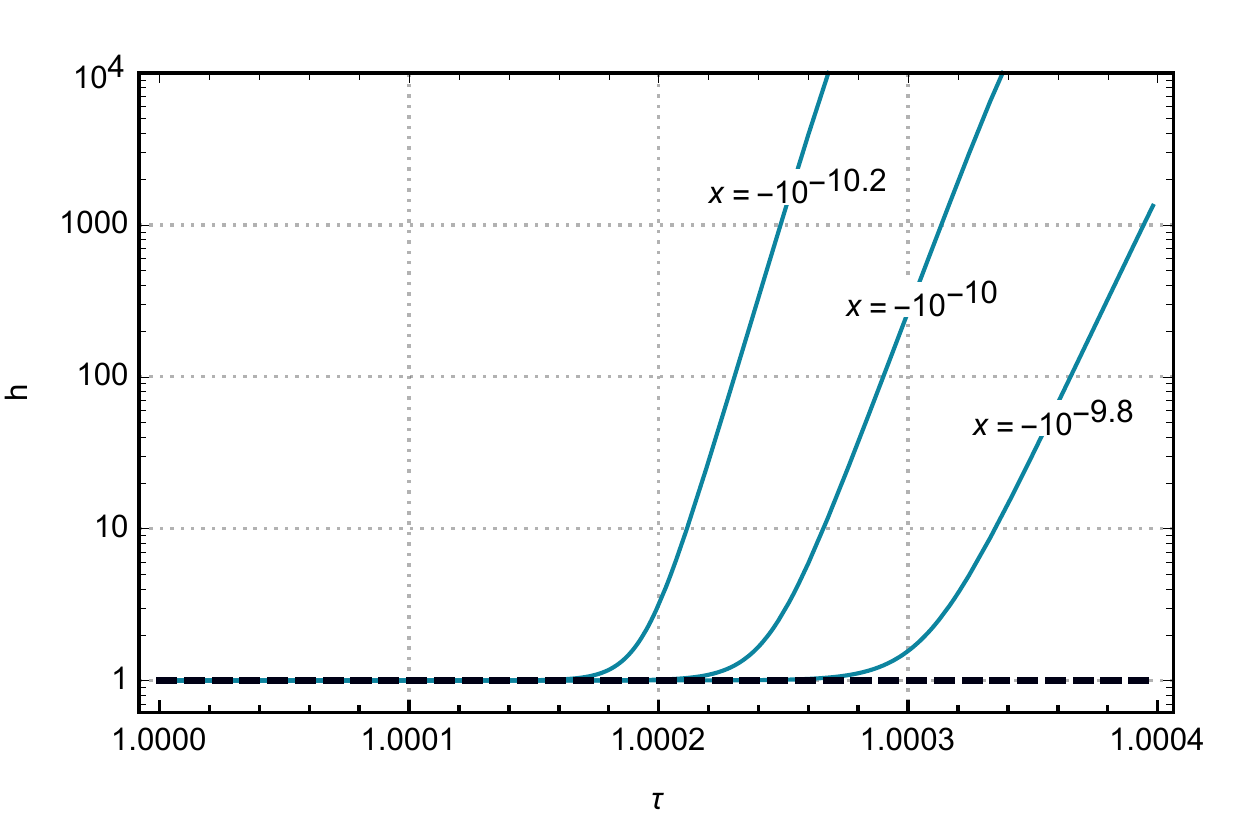}
	\includegraphics[scale=0.65]{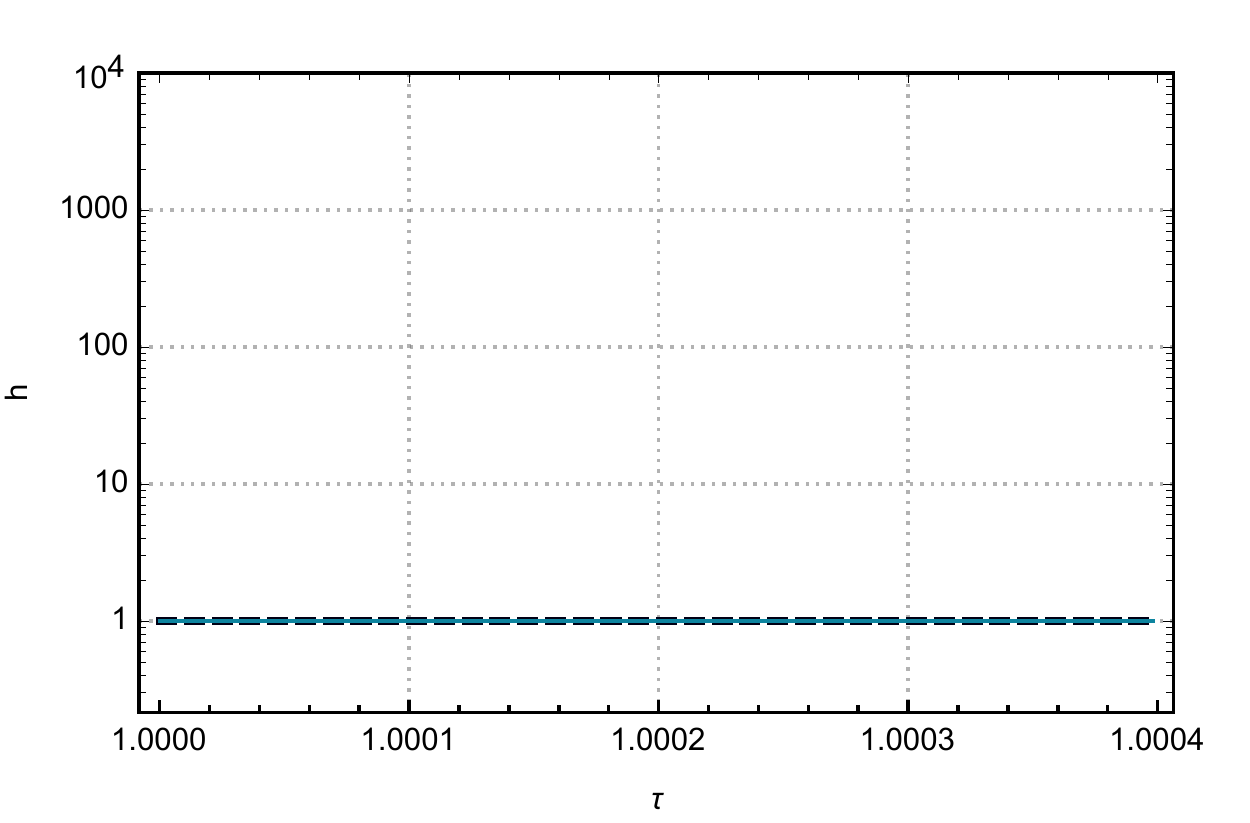}
	\caption{
We show the numerical solution of Eq.~(\ref{eq:hubble}) with the initial conditions and the 
higher-derivative couplings of Eq.~(\ref{eq:deSitter1-initial}).
These figures show that the dynamics of the dimensionless Hubble parameter $h( \tau)$ in a few normalization time $\tau$.
The dashed line shows the de Sitter solution $h( \tau)=1$ from the general relativity.
The right panel is $y =0$ whereas the tight panel is $y =0$.
The top panels shows that the solutions oscillate where 
where the top-to-bottom lines corresponds $x=10^{-10.0,-12.0,-14.0}$.
The bottom panels on the other hand shows the solutions exponentially grow 
where the top-to-bottom lines corresponds $x=-10^{-9.8,-10.0,-10.2}$.
We found out that the time-scale $\tau_I$ is $\tau_I \approx \left| x \right|^{1/2}\approx 10^{-5}$.}
	\label{fig:deSitter}
\end{figure*}

\begin{figure*}[ht]
	\centering
	\includegraphics[scale=0.65]{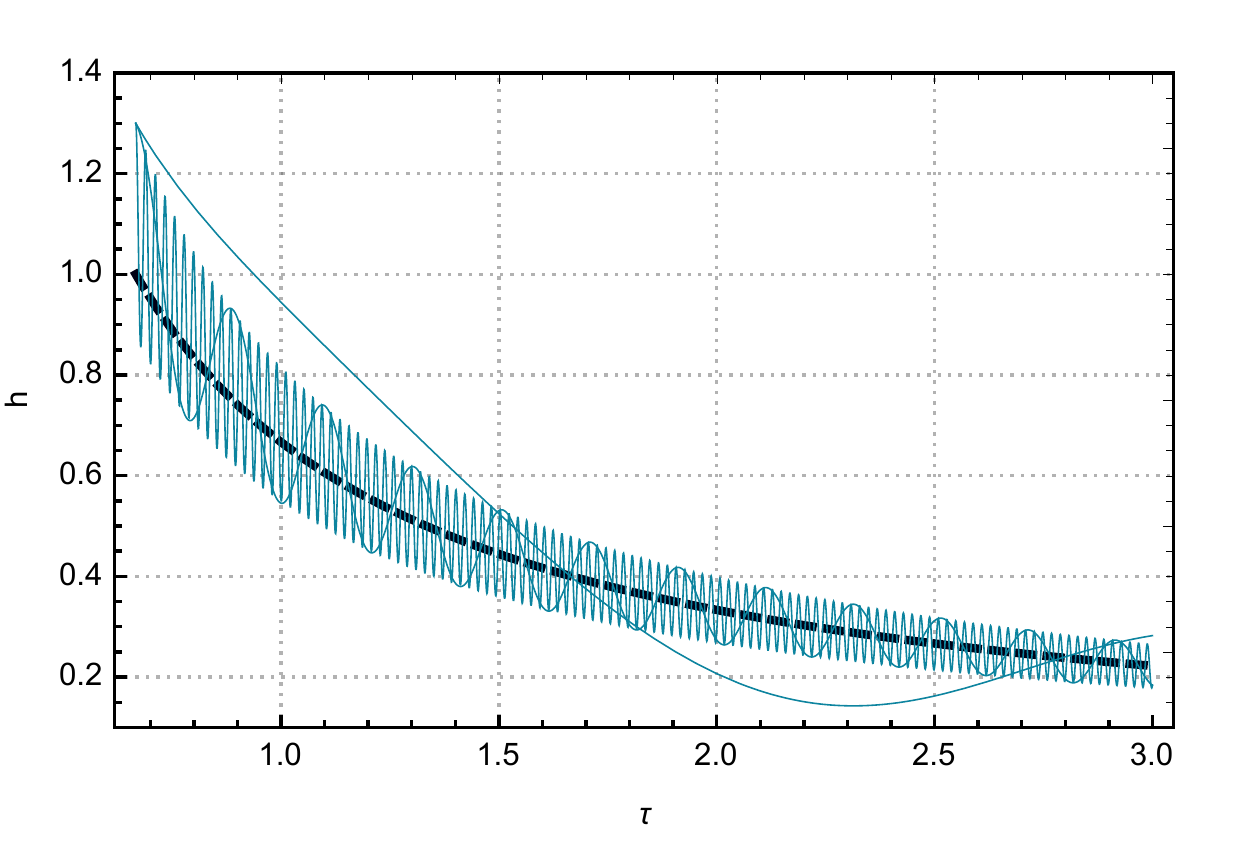}
	\includegraphics[scale=0.65]{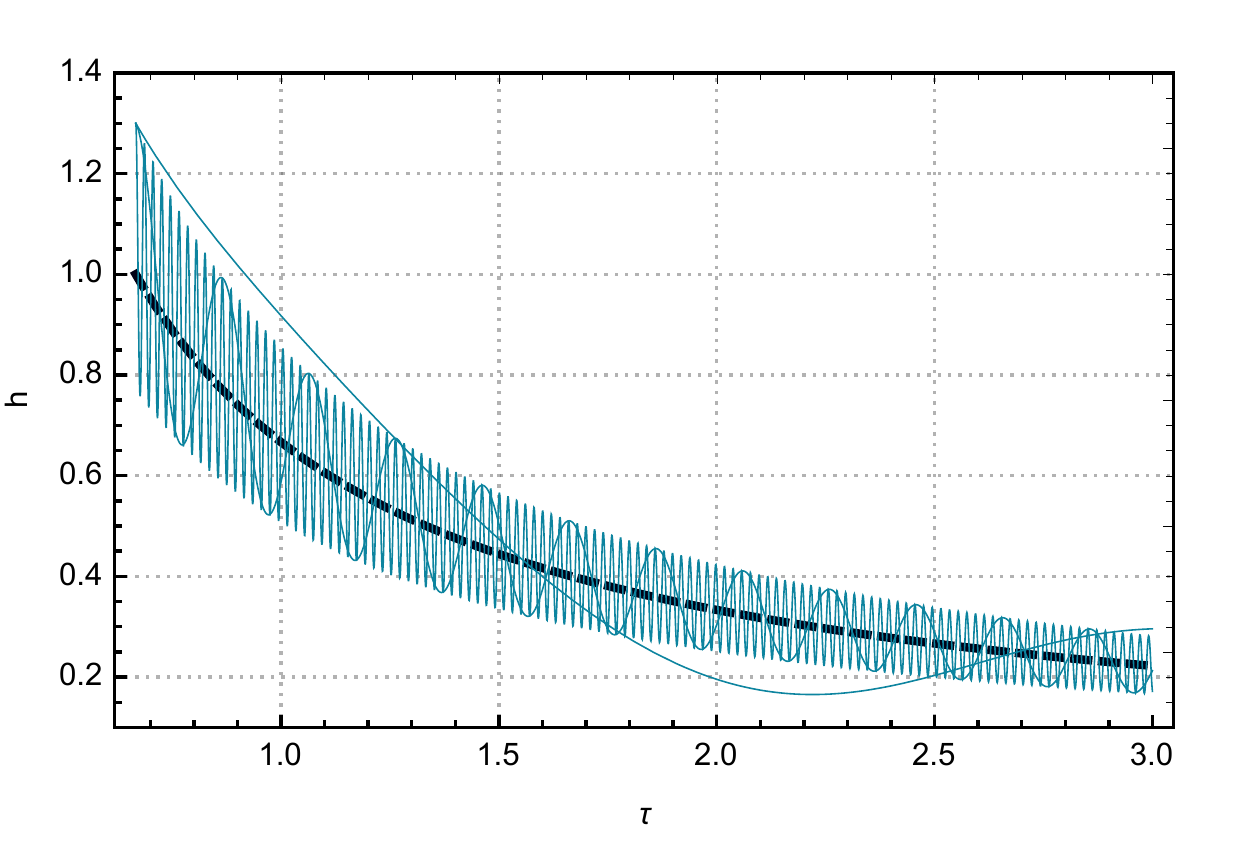}
	\includegraphics[scale=0.65]{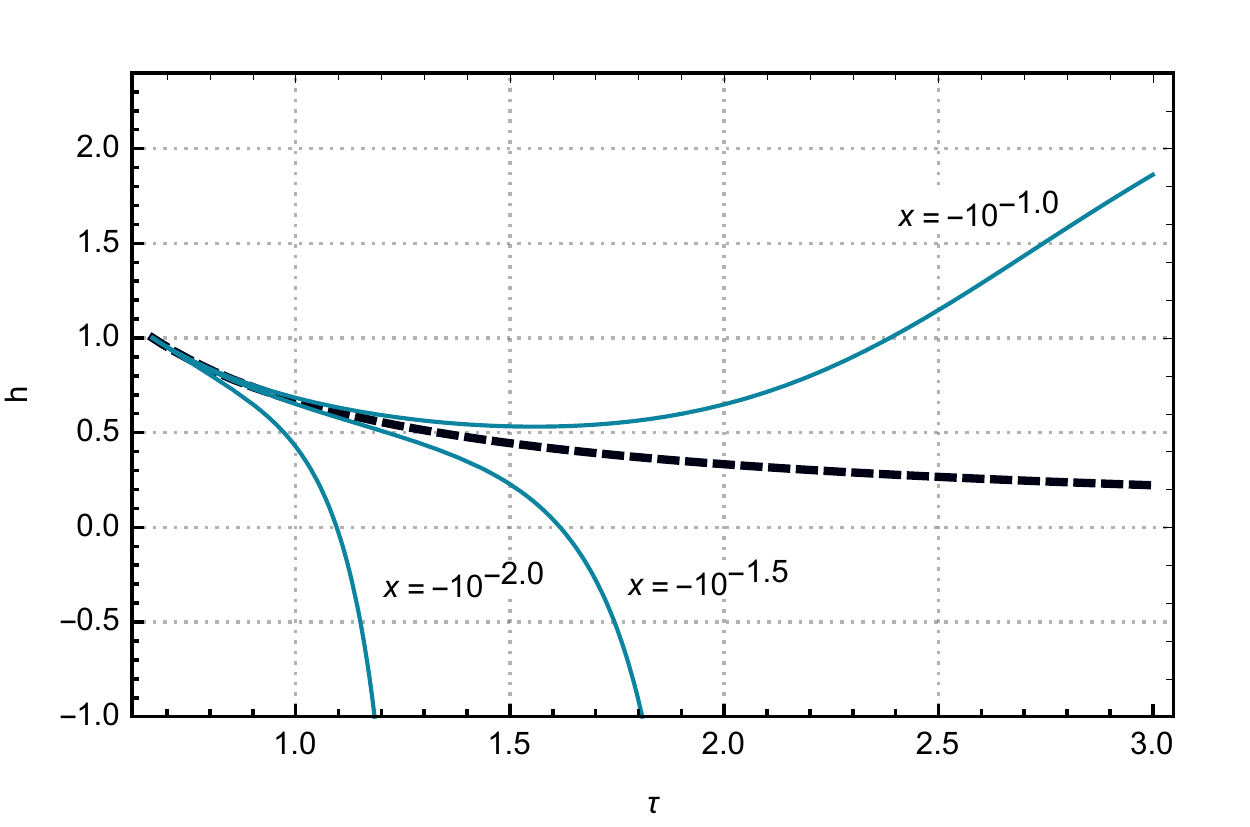}
	\includegraphics[scale=0.65]{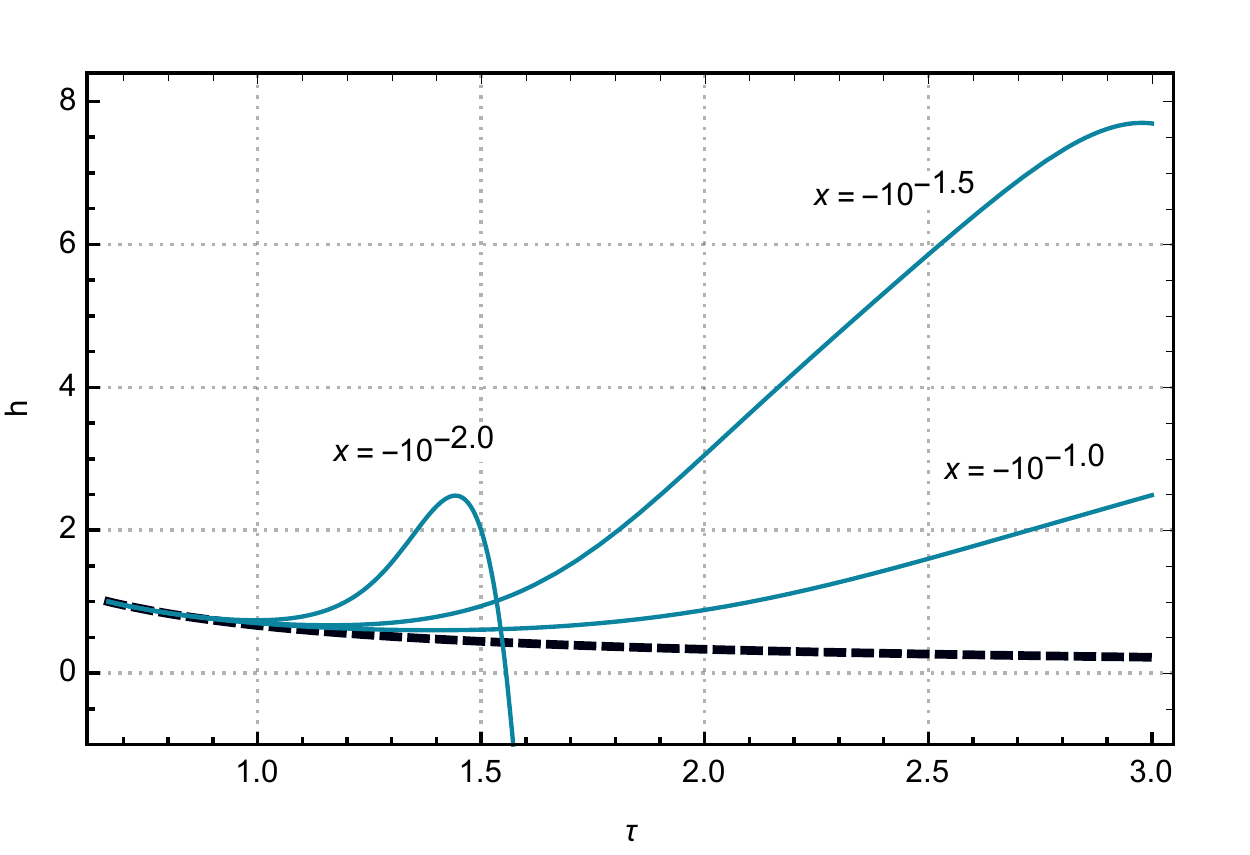}
	\caption{
We compare the numerical solution of Eq.~(\ref{eq:hubble}) with 
the conditions of Eq.~(\ref{eq:matter-initial})
and the standard solution $h( \tau)=1/2\tau$ from the general relativity
which is presented by the dashed line. 
The right panel is $y =0$ whereas the tight panel is $y =0$.
The top panels shows that the solutions oscillate for $x=10^{-1.0,-3.0,-5.0}$.
The bottom panels where the top-to-bottom lines corresponds $x=-10^{-1.0,-1.5,-2,0}$
shows the instabilities for the matter-dominated Universe
and the solutions are inconsistent with the standard general relativity.}
\label{fig:matter}
\end{figure*}

\subsection{Numerical analysis for FLRW spacetime instability}

Let us consider the de Sitter spacetime under the Hubble perturbation
and discuss the cosmological evolution.
Now using Eq.~(\ref{eq:semiclassical}) for the FLRW metric,
we obtain the semiclassical Friedmann equations,
\begin{align}\label{eq:Hubble-equation}
\begin{split}
{H^2}&= \frac{\Lambda}{3}
-{48\pi G_N }{\alpha_1}\left(6{ H }^{ 2 }\dot { H } +2H\ddot { H } -{ \dot { H }  }^{ 2 }\right)\\
&+{8\pi G_N }\alpha_3H^4+ \frac{8\pi { G }_{ N }}{3}\rho_{\rm m},
\end{split}
\end{align}
where $\rho_{\rm m}$ is the energy density of the classical matter 
and satisfies the covariant conservation law,
\begin{align}\label{eq:conservation-law}
\dot {\rho}_{\rm m}&=-3H\left( \rho_{\rm m}+P_{\rm m}\right)=-
3H\left( 1+\omega \right)\rho_{\rm m},
\end{align}
where $w=P/\rho$ is an equation-of-state parameter.
For the non-relativistic matter, radiation and cosmological constant 
one takes $w = 0,1/3,-1$, respectively.

We rewrite the above semiclassical equations in terms of 
the dimensionless parameters,
\begin{align}\label{eq:thubble}
\begin{split}
&h^{2}=-x\left(6h^{2}h'+2hh''-h'^{2} \right)+{y}h^{4}+z,\\
&z'=-
3h\left( 1+\omega \right)z\,, \end{split}
\end{align}
where these parameters are given by 
\begin{align}\label{eq:dimensionless}
\begin{split}
&\tau =H_0 t,\quad h=H/H_0,\\
&x={48\pi G_N }\alpha_1{ H }^{ 2 }_0,\quad
y={8\pi G_N }\alpha_3{ H }^{ 2 }_0,\\
&z= \Lambda/3{ H }^{ 2 }_0+{8\pi G_N }\rho_{\rm m}/3{ H }^{ 2 }_0 \,, 
\end{split}
\end{align}
where $H_0$ is the initial Hubble parameter.
We note the following relations of these parameters,
\begin{align}
\begin{split}
&H_0 \sim 10^{14 }\, {\rm GeV}, \ M_{\rm P} \sim 10^{18}\, {\rm GeV}, \ \alpha_{1,3}\sim 10^{-2}\\
&\ \Longrightarrow \ x,y\sim 10^{-10},\\
&H_0 \sim 10^{-42 }\, {\rm GeV}, \ M_{\rm P} \sim 10^{18}\, {\rm GeV}, \ \alpha_{1,3}\sim 10^{-2}\\
&\ \Longrightarrow \ x,y\sim 10^{-122},
\end{split}
\end{align}
where the formar and latter Hubble parameter  corresponds to an expample consistent with 
typical inflation and current universe, respectively~\cite{Aghanim:2018eyx}.
The dynamics of the dimensionless Hubble parameter $h$ with $w = -1$
for the spacetime is determined by,
\begin{align}\label{eq:hubble}
h^{2}=-x\left(6h^{2}h'+2hh''-h'^{2} \right)+{y}h^{4}+z\,,
\end{align}
where $z$ is constant and
prime express the derivative with respect to the dimensionless time $\tau$.
The suitable de Sitter initial conditions for Eq.~(\ref{eq:hubble}) are
\begin{equation}
\tau_0=1,\quad h_0=1,\quad h_0'=0,\quad  z_0=1.
\end{equation}

In order to investigate the de Sitter 
spacetime instabilities, we consider the numerical solutions 
of the de Sitter system starting at $\tau_0=1$
with various initial conditions and perturbations.
In Fig.\ref{fig:deSitter},
we present the numerical results for the dimensionless parameter $h(\tau)$ 
from Eq.~(\ref{eq:hubble}) with the following conditions,
\begin{align}
\begin{split}\label{eq:deSitter1-initial}
\textrm{Fig.\ref{fig:deSitter}}:&\ h_0=1+10^{-1.0},\  h_0'=0,\\
&\ x=10^{-10.0,-12.0,-14.0},\ y=0,10^{-10.0},\\
&\ h_0=1,\  h_0'=0,\\
&\ x=-10^{-9.8,-10.0,-10.2},\ y=0,10^{-10.0},
\end{split}
\end{align}
where we compare these results with the de Sitter solution 
$h( \tau)=1$ from the general relativity.
The top panels show that 
the Hubble perturbations in the de Sitter spacetime oscillate and
the oscillation time-scale becomes shorter for the small values of $\left| x \right|$. 
On the other hand, the bottom panels show that 
the de Sitter spacetime 
is destabilized in $\tau\approx \left| x \right|^{1/2}$ and the smallness of $\left| x \right|$ 
amplifies the instabilities.
The dynamics for $y>0$ and $y<0$
shows the similar results and 
it is found that $y$ is irreverent for the dynamics.

Next, we consider matter-dominated stage of the Universe with $w = 0$.
The natural initial conditions for the system are given by,
\begin{equation}
\tau_0=2/3,\quad h_0=1,\quad h_0'=-3/2,\quad z_0=1,
\end{equation}
where we take $z= {8\pi G_N }\rho_{\rm m}/3{ H }^{ 2 }_0$ and $\Lambda=0$.
In Fig.\ref{fig:matter}
we investigate the system of equations starting 
at $\tau_0=2/3$ with the following conditions,
\begin{align}
\begin{split}\label{eq:matter-initial}
\textrm{Fig.\ref{fig:matter}:}&\ h_0=1+0.3,\  h_0'=-3/2,\\
&\ x=10^{-1.0,-3.0,-5.0}\ y=10^{-1.0},\\
&\ h_0=1,\  h_0'=-3/2, \\
&\ x=-10^{-1.0,-1.5,-2,0},\ y=10^{-1.0},\\
\end{split}
\end{align}
where we compare them the standard cosmic solution $h( \tau)=2/3\tau$.
The top panels show that the matter-dominated Universe is unstable 
for the small perturbations but the solutions converge the general relativity. 
The bottom panels show that
the matter-dominated Universe is unstable for $\tau\approx \left| x \right|^{1/2}$
and inconsistent with the usual general relativity.
In this case, the instability time-scale $\tau_I$
should be larger than of order unity,
\begin{align}
\tau_I \approx \left| x \right|^{1/2}\gtrsim \mathcal{O}(1),
\end{align}
and thus using the current value of the Hubble parameter $H_0 \sim 10^{-42 }\, {\rm GeV}$,
we obtain the following constraint,
\begin{align}
\left|\alpha_{ 1 }\right|\gtrsim 10^{118},
\end{align}
which is consistent with Eq.~(\ref{eq:condition}).
Now, we confirmed the results of the previous subsection and found out that 
the FLRW spacetime is unstable under the perturbations for $\alpha_{ 1 }>0$ 
or the evolution for $\alpha_{ 1 }<0$.
Since the theoretically expected value of $\alpha_{ 1 }$ is given by Eq.~(\ref{eq:alpha}),
the above constraint is unacceptably large.

\begin{figure}[t]
\includegraphics[width=85
mm]{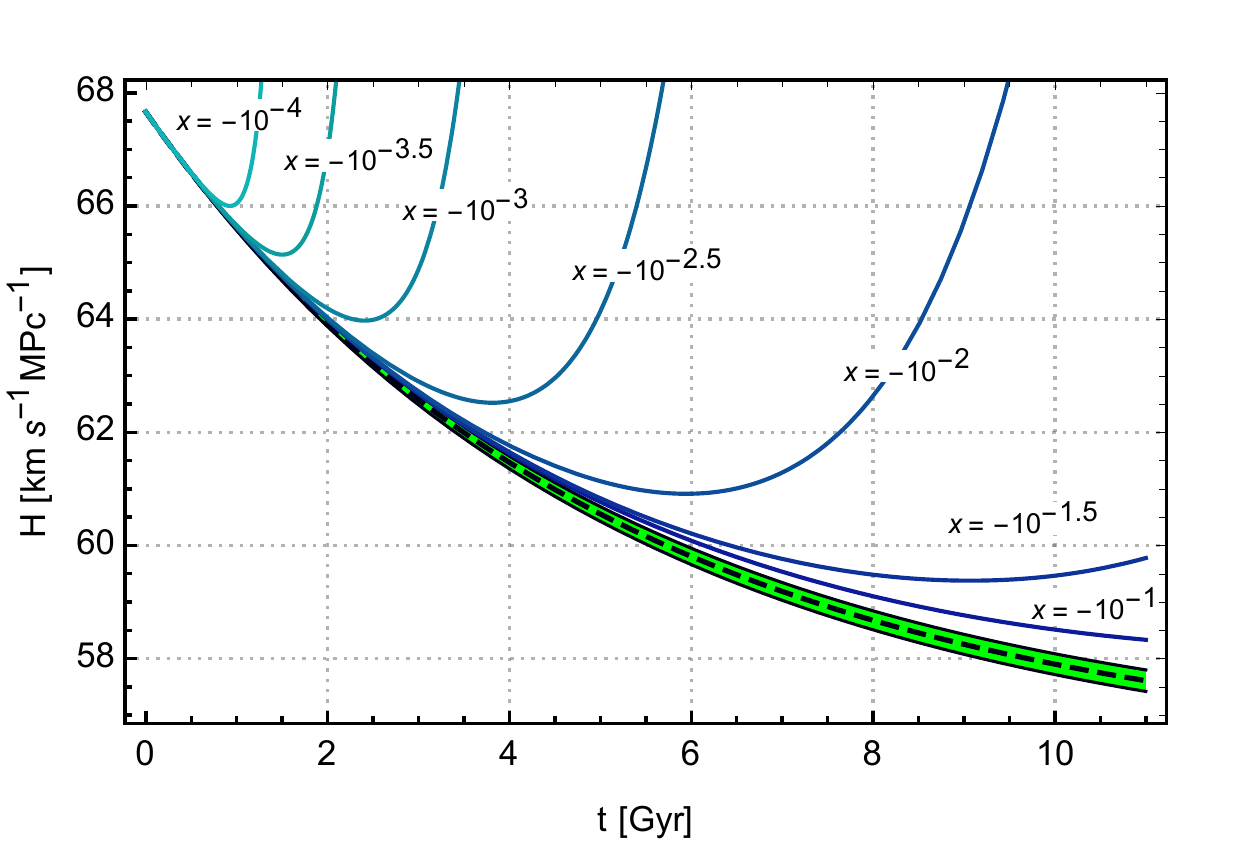}
\caption{
In this figure 
we compare standard result of the $\Lambda$CDM
and the numerical solutions of Eq.~(\ref{eq:thubble}) with 
the conditions of Eq.~(\ref{eq:LCDM-initial}) and Eq.~(\ref{eq:LCDM-condition}).
The dashed line corresponds to the central value of the 
Planck data~\cite{Aghanim:2018eyx} and the green line
expresses the allowed region of the  Planck data (Planck 2018, 
TT,TE,EE+lowE+lensing+BAO 68\% limits).
The semiclassical gravity or conformal anomaly expect
$\left| x \right|\approx 10^{-122}$ and they are not consistent with the observations.}
\label{fig:LCDM1}
\end{figure}

\subsection{Instability of the current Universe}
Finally, we consider cosmological evolution of the Universe
with the quantum back-reaction and 
derive a strict cosmological constraint for 
the semiclassical gravity.
Before considering the detail, let us review
the standard cosmological history of the Universe
based on $\Lambda$CDM with inflation.
First, the Universe proceeds inflation~\cite{Starobinsky:1980te,Guth:1980zm,
Sato:1980yn,Linde:1981mu,Albrecht:1982wi} (around $10^{-35}$ to $10^{-32}$ sec).
Second, the inflation ends and thermal radiation dominates up to the recombination 
with the cosmic microwave background (CMB) radiation
(around 379,000 years). After the radiation-dominated stage, 
the Universe is dominated by non-relativistic matters, and then
galaxies and clusters are gradually formed.
From about $9.8$ Gyrs, the expansion of the universe 
begins to accelerate via unknown dark energy. The age of the current Universe 
is about $13.8$ Gyrs and the present value of the Hubble parameter 
is $H_0 \approx 67.7\, {\rm km/s.MPc.}$

We consider the semiclassical Friedmann equations of Eq.~(\ref{eq:Hubble-equation})
for the conditions of $\Lambda$CDM,
\begin{align}\label{eq:LCDM}
\begin{split}
&H^2(t_0)=
{H_0^2}\left[ \Omega_{\Lambda,0}+\Omega_{{\rm m},0}
+\Omega_{{\rm r},0}\right],\\
&\qquad \Omega_{\Lambda,0}+\Omega_{{\rm m},0}
+\Omega_{{\rm r},0}\approx 1, 
\end{split}
\end{align}
where $\Omega_{\Lambda,0}$ is the density parameter 
for the cosmological constant and
$\Omega_{{\rm m},0}$, $\Omega_{{\rm r},0}$ are the current 
density values of noh-reativistc matters 
including dark matter, and radiation.
From the recent Planck data~\cite{Aghanim:2018eyx} (Planck 2018, 
TT,TE,EE+lowE+lensing+BAO 68\% limits) we have,
\begin{align}\label{eq:LCDM-initial}
\begin{split}
&H_0\approx  67.66 \pm 0.42\, [{\rm km\, s^{-1}\, MPc^{-1}}]\\
&\ \quad \approx  7.25 \times 10^{-2}\, [{\rm Gyr^{-1}}],\\
&\Omega_{\Lambda,0}=0.6889 \pm 0.0056,\\
&\Omega_{{\rm m},0}= 0.3111 \pm 0.0056 , 
\end{split}
\end{align}
where $(G_N)^{1/2}=10^{-59}\ {\rm Gyr.}$ 
For simplicity, we rewrite the semiclassical Friedmann equations as follows,
\begin{align}\label{eq:thubble}
\begin{split}
&h^{2}=-x\left(6h^{2}h'+2hh''-h'^{2} \right)+yh^{4}+\lambda_0+z,\\
&z'=-
3h z\,, \end{split}
\end{align}
where the dimensionless parameters are given by 
\begin{align}\label{eq:dimensionless}
\begin{split}
&\tau =H_0 t,\quad h=H/H_0,\\
&x={48\pi G_N }\alpha_1{ H }^{ 2 }_0,\quad
y={8\pi G_N }\alpha_3{ H }^{ 2 }_0,\\
&\lambda_0= \Lambda/3{ H }^{ 2 }_0,\quad z= {8\pi G_N }\rho_{\rm m}/3{ H }^{ 2 }_0 \,. 
\end{split}
\end{align}
In Fig.\ref{fig:LCDM1} we assume the following initial conditions
and various couplings based on the $\Lambda$CDM,
\begin{align}\label{eq:LCDM-condition}
\begin{split}
\textrm{Fig.\ref{fig:LCDM1}:}&
\ x=-10^{-1.0,-2.0,-3.0,-4.0},\ y=0,\\
&\ h_0=1,\quad h_0'=-3/2\, \Omega_{{\rm m},0},\\
&\ \lambda_0=\Omega_{\Lambda,0}
\quad z_0=\Omega_{{\rm m},0}.
\end{split}
\end{align}
The Fig.\ref{fig:LCDM1} shows the spacetime instabilities for the small $\left| x \right|$ 
in a short time and the corresponding spacetime solutions 
are inconsistent with the future or current 
evolution of the Universe 
unless one takes $\left| x \right|^{1/2}\approx \mathcal{O}(1)$.
However, the semiclassical gravity expects the extremal small value 
$\left| x \right|^{1/2}\approx 10^{-61}$ for the current Universe and 
that is not consistent with the observations.

The spacetime instability of the semiclassical gravity 
is certainly a serious problem and the 
solutions are not be consistent with the cosmological observation.
It has been argued that the semiclassical solutions must be
given by the truncating perturbative expansions~\cite{
Simon:1991bm,Parker:1993dk} and the quantum higher derivative corrections 
can be regarded as small perturbations from the classical solution.
However, this procedure is ad hoc approach for the semiclassical equations
much below the Planck scale and ineffective near the Planck regime.
That is a problem when one consider large $N$ expansion where the semiclassical gravity could 
be adequate to describe Planckian phenomena due to 
the suppression of the graviton loops~\cite{Hartle:1981zt}.
Also the Euclidean formulation of quantum gravity  
imposes the boundary condition and the curvature instability or runaway solutions
might be removed~\cite{Hawking:2001yt}.
However, it is not clear how to 
handle the quantum energy momentum tensor $\left< { T }_{ \mu\nu }\right>$ 
in these procedures and the problem of the quantum instability is still left open.

\section{Conclusion }
\label{sec:discussion}
Semiclassical gravity describe the interactions between
classical gravity and quantum matters, and 
the quantum back-reaction is formally defined as the 
higher-derivative curvatures. These induce instabilities of the classic solutions 
and Refs~\cite{Horowitz:1978fq,Horowitz:1980fj,Hartle:1981zt,RandjbarDaemi:1981wd,Jordan:1987wd,Suen:1988uf,Suen:1989bg,Anderson:2002fk} presented 
that the Minkowski spacetime is unstable under small perturbations.
The spacetime instability was seen as a serious problem in semiclassical gravity.
However, it has not been discussed whether the semiclassical instabilities are inconsistent in our Universe.

In this paper we have shown that the homogenous and isotropic FLRW Universe interacting 
with quantum matter fields are unstable under small perturbations or the evolutions.
We have analytically and numerically demonstrated
that the homogenous and isotropic cosmological solutions either grow exponentially 
or oscillate even in the Planckian time $t_{\rm I}=\alpha_{ 1 }10^{-43}\ {\rm sec}$. For $\alpha_{ 1 }>0$, the curvature 
perturbations oscillate rapidly and would emit the Planck energy photons~\cite{Horowitz:1978fq}
which is unacceptable for the observed Universe.
On the other hand, for $\alpha_{ 1 }<0$, the evolution of the curvature perturbation 
leads to the Planckian curvature or singularity. 
These instabilities induce a catastrophe
unless one takes extremal values of the gravitational 
couplings or fundamental particle species $\left|\alpha_{ 1 }\right|\gtrsim 10^{118}$.
We have also confirmed these results based on the cosmological evolution
by comparing $\Lambda$CDM and 
the semiclassical Einstein solutions using the Planck data and it is found that
these solutions of the semiclassical gravity 
including conformal anomaly are not consistent with cosmological observations.

\medskip
\para{Acknowledgements} 
NW would like to thank Shingo Kukita for the valuable discussions.

\nocite{}
\bibliography{Reference}
\end{document}